\documentclass[aps,pra,a4paper,reprint,twocolumn,superscriptaddress,nofootinbib,floatfix]{revtex4-1}
\pdfoutput=1
\usepackage{amsmath}
\usepackage{amssymb}
\usepackage{verbatim}
\usepackage{graphicx}
\usepackage{tabularx}
\usepackage{xcolor}
\usepackage{mathtools}
\usepackage{bbold}
\usepackage{blkarray}
\usepackage{appendix}
\usepackage[inline]{enumitem}
\usepackage{latexsym}
\usepackage{array}
\usepackage{colortbl}
\usepackage{bm} 
\usepackage{physics}
\usepackage{dsfont} 
\usepackage{hyperref}
\usepackage{nicefrac}
\usepackage{soul} 
\hypersetup{colorlinks=true,        
	linkcolor=black,         
	citecolor=black,        
	filecolor=black,      
	urlcolor=black }

\newcommand\blfootnote[1]{%
	\begingroup
	\renewcommand\thefootnote{}\footnote{#1}%
	\addtocounter{footnote}{-1}%
	\endgroup
}

\newcommand{\methods}{Appendix}
\newcommand{\si}{Appendix}

\newcommand{\plaqop}{\hat{\square}}
\renewcommand\vec{\mathbf}

\newcommand{\figref}[1]{Fig.~\ref{#1}}
\newcommand{\figrefp}[2]{Fig.~\ref{#1}{#2}}

\newcommand{\uibkexp}{\affiliation{Universit\"at Innsbruck, Institut f\"ur Experimentalphysik, Technikerstraße 25a, Innsbruck, Austria}}
\newcommand{\uibkth}{\affiliation{Universit\"at Innsbruck, Institut f\"ur Theoretische Physik, Technikerstraße 21a,  Innsbruck, Austria}}
\newcommand{\iqoqi}{\affiliation{Institute for Quantum Optics and Quantum Information of the Austrian Academy of Sciences,  Technikerstraße 21a, Innsbruck, Austria}}
\newcommand{\aqt}{\affiliation{Alpine Quantum Technologies GmbH, Innsbruck, Austria}}
\newcommand{\waterloo}{\affiliation{Institute for Quantum Computing, University of Waterloo, Waterloo, Ontario, N2L 3G1, Canada}
	\affiliation{Department of Physics \& Astronomy, University of Waterloo, Waterloo, Ontario, N2L 3G1, Canada}
}
\newcommand{\exeter}{\affiliation{Department of Physics and Astronomy, University of Exeter, Stocker Road, Exeter EX4 4QL, United Kingdom}}
\newcommand{\ulm}{\affiliation{Institut für Theoretische Physik und IQST, Universität Ulm, Albert-Einstein-Allee 11, D-89069 Ulm, Germany}}
\newcommand{\perimeter}{\affiliation{Perimeter Institute for Theoretical Physics, Waterloo, Ontario, N2L 2Y5, Canada}}

\begin{document}
	
	\title{Simulating 2D lattice gauge theories on a qudit quantum computer}
	\author{Michael Meth}
	\uibkexp
	\author{Jan F. Haase}
	\waterloo
	\ulm
	\author{Jinglei Zhang}
	\waterloo
	\author{Claire Edmunds}
	\uibkexp
	\author{Lukas Postler}
	\uibkexp
	\author{Andrew J. Jena}
	\waterloo
	\author{Alex Steiner}
	\uibkexp
	\author{Luca Dellantonio}
	\waterloo
	\exeter
	\author{Rainer Blatt}
	\uibkexp
	\iqoqi
	\aqt
	\author{Peter Zoller}
	\uibkth
	\iqoqi
	\author{Thomas Monz}
	\uibkexp
	\aqt
	\author{Philipp Schindler}
	\uibkexp
	\author{Christine Muschik$^*$}
	\waterloo
	\perimeter
	\author{Martin Ringbauer$^*$}
	\uibkexp
	\blfootnote{* These authors jointly supervised the work.}
	
	\begin{abstract}
		\noindent 
		Particle physics underpins our understanding of the world at a fundamental level by describing the interplay of matter and forces through gauge theories. Yet, despite their unmatched success, the intrinsic quantum nature of gauge theories makes important problem classes notoriously difficult for classical computational techniques~\cite{Banuls:2019rao,creutz1983monte}. A promising way to overcome these roadblocks is offered by quantum computers, which are based on the same laws that make the classical computations so difficult~\cite{feynmanQS,banuls2019simulating,Dalmonte_2016,funcke2023review,dimeglio2023quantum,beck2023quantum}. Here we demonstrate two essential requirements for gauge theory calculations with quantum computers. Firstly, we perform a quantum computation of the properties of the basic building block of two-dimensional lattice quantum electrodynamics, involving both gauge fields and matter ~\cite{paulson_simulating_2021,zohar_quantum_2021}. Secondly, we show how to refine the gauge field discretization beyond its minimal representation.
			These computations are made possible by the use of a trapped-ion qudit quantum processor~\cite{Ringbauer2022A-universal}, where quantum information is encoded in $d$ different states per ion, rather than in two states as in qubits. Such qudits are ideally suited for describing gauge fields, which are naturally high-dimensional, leading to a dramatic reduction in the register size and circuit complexity. Using a variational quantum eigensolver~\cite{Farhi2014, PeruzzoVQE, OMalley2016, McClean2016, Moll2018} we prepare the ground state of the model. We then observe the effect of dynamical matter on quantized magnetic fields and show how to seamlessly observe the effect of different gauge field truncations by controlling the qudit dimension. Our results open the door for hardware-efficient quantum simulations with qudits in near-term quantum devices.

	\end{abstract}
	
	\maketitle
	
	\section{Introduction}
	\label{sec:introduction}
	
	Computing today is almost exclusively based on binary information encoding. This holds true for classical computers operating with bits, as well as for the emerging area of quantum computing that uses qubits to exploit quantum superposition and entanglement for information processing. 
	However, quantum systems underpinning today’s quantum computers offer the possibility to process information in several different energy levels~\cite{Ahn2000, Godfrin2017, Anderson2015, Morvan2020, Senko2015, Wang2018, Chen_2021}, so-called qudits. A key to unlocking the potential of this approach, and to realizing qudit algorithms~\cite{wang_qudits_2020} in practice is the availability of programmable, high-fidelity qudit entangling gates. We realize this capability in a linear ion-trap quantum processor with all-to-all connectivity~\cite{Ringbauer2022A-universal} by extending qubit entangling gates~\cite{Cirac1995, Kuzmin2020, Meth2022} to mixed-dimensional qudit systems. These resources open up exciting avenues for native quantum simulation of $d$-level systems (e.g.\ in chemistry~\cite{Cao2019Quantum, MacDonell2020, maskara_quantum_2023} or condensed matter physics~\cite{Haldane1983, Sawaya2020}) with smaller registers and reduced gate depth compared to a qubit approach.
	
	A natural application for qudit quantum hardware is lattice gauge theory (LGT) calculations, where qudits naturally represent high-dimensional gauge fields. Gauge theories are the backbone of the standard model of particle physics. Studying them on a lattice through computer simulations has been key in the quest for a more complete understanding of the phenomenology within the standard model and for discovering physics beyond it. Yet, despite the tremendous success of LGT classical simulations~\cite{aoki_flag_2022}, this endeavor is increasingly hindered by the fact that important problem classes, such as real-time evolutions and problems involving high matter densities, are plagued by sign problems~\cite{Gattringer:2016kco}, which are believed to be classically intractable~\cite{Troyer:2004ge, jordan_bqp-completeness_2018}. Quantum computations, on the other hand, are by design not affected by sign problems and thus offer a unique scientific opportunity for advancing the frontier of gauge theory simulations (see~\cite{dimeglio2023quantum,beck2023quantum,bauer_quantum_2023} for an in-depth discussion and \ref{Sec:Supp:Advantage} for key points). While LGT quantum simulations for particle physics have seen impressive advances, experimental demonstrations were limited to either one spatial dimension (1D), or targeted theories beyond 1D where either gauge fields or matter are trivial~\cite{klco20192,Trailhead,Rahman2021,Ciavarella2022,Rahman2022A,Rahman2022B,Ciavarella2023}.

Here we address two major challenges in quantum computing for gauge theory calculations: The realization of LGT quantum computations beyond one spatial dimension (1D) including both gauge fields and matter and the ability to control the gauge field dimension. Both of these essential ingredients demand the efficient representation of the formally infinite-dimensional gauge fields on quantum computers, which requires discretization and truncation to a finite number of levels~\cite{Digitization2019, PhysRevD.100.114501, paper1, davoudi_search_2021,Trailhead, Bauer2023, jakobs_canonical_2023}. Importantly, however, truncated gauge fields must remain sufficiently high-dimensional to capture the relevant physics, which is most naturally achieved by encoding them into qudits~\cite{yang_analog_2016, Mil:2019pbt, gonzalez-cuadra_hardware_2022, zache2023}.

Specifically, we consider quantum electrodynamics in two spatial dimensions (2D-QED) and simulate the basic building block of the lattice---a single plaquette---on a qudit quantum computer~\cite{Ringbauer2022A-universal}. 
We observe the ground state plaquette expectation value~\cite{creutz1983monte}, which is a central quantity in LGT calculations related to magnetic fields that are a defining feature of 2D and 3D physics and have no analog in 1D. The plaquette expectation value is also relevant for the so-called running coupling~\cite{aoki2017review,aoki2020flag} in gauge theories, which is absent in 1D-QED. Notably, our approach is directly adaptable to digital quantum simulations of real-time dynamics, offering an intriguing perspective for the quantum simulations of LGT dynamics. In all demonstrated cases, our qudit-encodings with high-fidelity entangling gates natively allow for gate sequences with fewer entangling gates, smaller register sizes, and improved simulation accuracy compared to conventional qubit devices. Providing an order of magnitude improvement in circuit complexity already in the simplest instance, our work highlights a new path towards highly efficient quantum simulations of gauge theories and beyond.

\section{LGT simulations with qudits}
\label{sec:hardware}
\begin{figure*}[th!]
\centering
\includegraphics[width=0.98\textwidth]{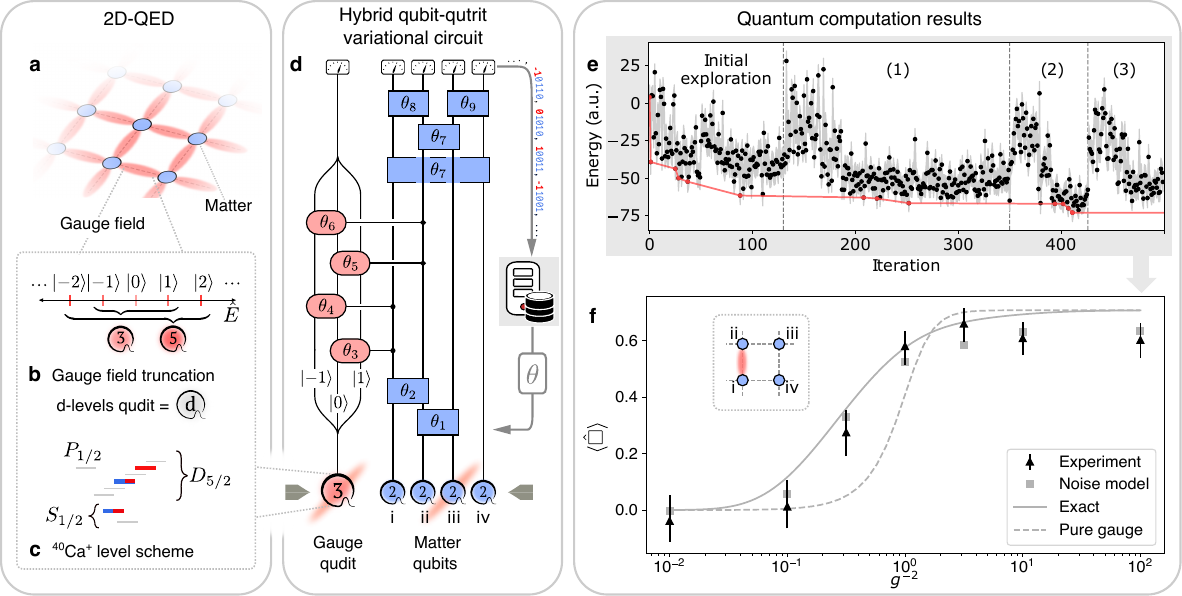}
\caption{\textbf{Simulating two-dimensional lattice QED with matter fields.} \textbf{a}, The lattice of 2D-QED is comprised of vertices containing matter particles (blue) connected by links carrying the associated gauge field (red). The local gauge field at a particular link possesses an infinite, yet discrete spectrum that can be simulated using a truncated representation. \textbf{b}, The truncated gauge fields are naturally represented by $d$-level qudits. \textbf{c}, In our experiment these qudits (red) are encoded in different Zeeman levels of the S$_{1/2}$ ground state and D$_{5/2}$ excited state of trapped $^{40}\mathrm{Ca}^+$ ions. Matter particles, on the other hand, are faithfully represented by qubits (blue), marking their presence or absence. \textbf{d}, Our variational ansatz for simulating a single plaquette with open boundary conditions is thus a mixed-dimensional qubit-qudit quantum circuit, see Sec.~\ref{sec:open}. Employing Gauss' law to eliminate three of the four gauge fields, we truncate the remaining field to contain at most one energy quantum, hence our quantum register contains one qutrit (red) for the gauge field and four qubits (blue) representing the matter vertices of the plaquette. A fanning-out of the qudit circuit line is used to illustrate the coupling of the qudit entangling gates to certain subspaces, see also \methods~\ref{sec:circuit_gates}. Note that the all-to-all connectivity of the ion trap quantum processor enables a highly efficient circuit realization. During the variational optimization, a classical optimizer varies the gate angles $\theta_j$ in order to minimize the energy of the produced quantum state, as measured by the quantum computer. \textbf{e}, An exemplary optimization run for $\Omega=5$, $m=0.1$, and $g^{-2}=10^2$. As a guide for the eye, the red line highlights the current lowest energy found by the algorithm. The first block of 130 evaluations explores generally the variational landscape. Subsequent blocks of up to 300 evaluations ((1), (2), (3), ... not all shown) optimize each for decreasing values of the coupling starting from  $g^{-2}=10^{2}$. During the whole optimization process, measurement data acquired at each iteration is used also to further explore the energy landscape for previously analysed values of the coupling. See \methods~\ref{sec:vqe} for details.
One standard deviation of statistical uncertainty is given by the shaded region. Once an optimal set of parameters $\vec{\theta}^*$ is found, the properties of the so-created approximate (ground) state can be studied. 
\textbf{f}, The expectation value of the plaquette operator $\langle\plaqop\rangle$ as defined in the main text. The triangular data points are measured on the VQE optimized state, with error bars representing one standard deviation of statistical noise. The squares are obtained from a numerical simulation of an ideal VQE, with an experimentally motivated noise model applied to the final state, see \si~\ref{sec:noise_model}. The line is obtained from a classical calculation of the exact ground state, using exact diagonalization. The dashed line is obtained from the corresponding pure gauge model $g^2 \hat{H}_E + \frac{1}{g^2} \hat{H}_B$: the presence of dynamical matter noticeably affects the slope of the plaquette expectation value when varying the coupling.}
\label{fig:1}
\end{figure*}

%
We simulate lattice QED, defined on a two-dimensional discretization of space, where matter (electrons and positrons) reside on the sites of the lattice, and gauge bosons (photons) on the links, see Fig.~\ref{fig:1}a. Matter is described by fermionic field operators, where we use a staggered formulation~\cite{kogut1975hamiltonian} in which lattice sites can either be in the vacuum state or host electrons (positrons) residing on even (odd) lattice sites, as depicted in \figref{fig:encoding_si} in the \methods.
The gauge fields residing on the links between each pair of sites are each described by electric field operators 
$ \hat{E}$ that possess an infinite, but discrete, spectrum $\hat{E} \ket{E} = E \ket{E}$, where $E = 0,\, \pm1,\,\pm2,\dots$
The total Hamiltonian is then given by the sum~\cite{kogut1975hamiltonian} 
\begin{eqnarray}\label{eq:Hamiltonian}
\hat{H} = g^2\hat{H}_E +\frac{1}{g^2} \hat{H}_B + m\hat{H}_m + \Omega\hat{H}_k,  
\end{eqnarray}
which describes the free electric (E), magnetic (B), and matter (m) field energy and the kinetic energy term (k) responsible for pair creation processes, see Eqs.~\eqref{eq:Hamiltonian_terms} in the \methods.

The bare coupling strength $g$ is determined by the charge of the elementary particles with bare mass $m$. Both parameters enter into the energy cost associated with the creation of an electron-positron pair and the associated electromagnetic field. The rate of these pair- and field-creation processes is characterized by $\Omega$. We employed here the Kogut-Susskind Hamiltonian formulation~\cite{kogut1975hamiltonian} in natural units $\hbar=c=1$, and with lattice spacing $a=1$.

Importantly, not all quantum states in the considered Hilbert space are physical. In particular, the gauge field and charge configurations of physical states $|\Psi_{\text{phys}}\rangle$ have to fulfill Gauss' law at each site. Here, the familiar law from classical electrodynamics
$\nabla \vec{E}(\vec{r})-\rho(\vec{r})=0$, where $\rho(\vec{r})$ is the charge density at point $\vec{r}$, takes the form $\hat{G}_{\vec{n}}|\Psi_{\text{phys}}\rangle=0$, with the Gauss operator $\hat{G}_{\vec{n}}$ at lattice site $\vec n$ given in Eq.~\eqref{eq:gauss_operator} in the \methods.

To observe 2D effects in this model, we study the local plaquette operator $\plaqop = - \frac{1}{V}\hat{H}_B$, where $V$ is the number of plaquettes. The plaquette operator involves four gauge fields forming a closed loop along a single plaquette, see Eq.~\eqref{eq:P_op} and \figrefp{fig:encoding_si}{b} in the \methods. Since this observable is related to the curl of the vector potential, it is defined in at least two spatial dimensions and has no analogue in 1D-QED. The dependence of the plaquette ground state (i.e.~vacuum) expectation value $\langle \plaqop\rangle$ on $g^{-2}$ can be related to the running of the coupling~\cite{Clemente2022}, which is a fundamental feature of gauge theories in particle physics that captures the dependence of the charge on the distance (energy scale) on which it is probed.
%

Quantum simulations of 2D-LGTs face the difficulty of finding an adequate representation for the gauge field operators $\hat{E}$. While the fermionic field can be straightforwardly transformed to qubits~\cite{jordan1928pauli}, whose states either represent the presence or absence of a particle (see \methods~\ref{sec:Methods_EncodedHamiltonian} and \figrefp{fig:encoding_si}{c}), the gauge field requires a truncation of its spectrum and a description containing at least three quantum states, representing positive, zero, and negative flux values. 

In principle, such a representation could be constructed from qubits. However, in practice, using qubits drastically increases the quantum register size and immediately results in complex many-body interactions~\cite{paulson_simulating_2021}, see \si~\ref{Sec:Supp:Advantage} for details. For example, encoding $d$-level gauge fields requires at least $\lceil\mathrm{log}_2(d)\rceil$ qubits and even the application of local gauge field operators involves $\mathcal{O}(d^2)$ two-qubit gates~\cite{Sawaya2020}. Gauss' law requires that the creation or annihilation of particles occurs with the corresponding change in flux, which necessitates the application of gauge-field rising or lowering operators that are controlled by the state of the matter configuration. Implementing such controlled gauge-field operations requires an even higher qubit gate count than the local gauge-field operators. 
We circumvent this issue by representing each gauge field directly with a qudit system, containing exactly as many levels as required for the chosen truncation. As a consequence, local operations on the gauge fields remain local in the quantum computer, and the coupling between a matter site and a gauge field is realized as a two-body qubit-qudit interaction. We achieve an efficient implementation of these interactions through the use of explicit entanglement between the ion's internal state and the common motional mode in the spirit of the Cirac-Zoller gate~\cite{Cirac1995}. Conditional on the state of one ion (regardless of qudit dimension) a phonon is injected into the motional mode. If the phonon is present, a local operation is performed on the second ion, otherwise not. In the end, the phonon is deterministically removed from the motional mode and the internal states of the two ions are left entangled, see \methods~\ref{sec:experimetal_details} for details. 
We realize the qubits for the matter fields and qudits for the gauge fields within the $4S_{1/2}$ ground state and the $3D_{5/2}$ excited state manifolds of trapped $^{40}\mathrm{Ca}^+$ ions~\cite{Ringbauer2022A-universal}, and thereby demonstrate quantum computations with tailored mixed-dimensional quantum systems. As shown in \figrefp{fig:1}{d} and \methods~\ref{sec:vqe}, we perform a variational ground-state search, using a suitable ansatz in the form of a quantum circuit with gates that are parameterized by a set $\vec{\theta}$ (see Sec.~\ref{sec:open} below). A classical optimizer then varies the gate parameters to minimize the energy $\langle \hat{H} \rangle$ of the prepared states, which serves here as a cost function and is measured by the quantum computer. 
This optimization loop is repeated until the energy is minimized and one obtains a parameter set $\vec{\theta}^*$, such that the ansatz approximates the ground state of $\hat{H}$.

\section{Simulating gauge fields and matter}
\label{sec:open}

In our first experiment, we study 2D-QED on a lattice with open boundary conditions, crucially including both matter and gauge fields. In contrast to previous experiments (simulating LGTs in 1D or without matter) we observe the effects of virtual pair-creation, electromagnetic fields, and their interplay~\cite{paulson_simulating_2021} by studying the ground state expectation value of the plaquette operator $\langle\plaqop \rangle=-\frac{1}{V} \langle \hat{H}_B\rangle$.
Specifically, we consider the basic building block of the two-dimensional lattice, i.e.\ a single plaquette with open boundary conditions, consisting of four matter sites and four gauge fields, see inset in \figrefp{fig:1}{f}. These gauge and matter degrees of freedom are constrained by Gauss' law in Eq.~\eqref{eq:gausslaw} at each vertex. These constraints can be encoded explicitly into the Hamiltonian by eliminating redundant gauge degrees of freedom as shown in \methods~\ref{sec:Methods_EncodedHamiltonian}. As a result, the resource requirements for our quantum computation are significantly reduced, while at the same time, the simulated states are guaranteed to be physical, i.e.~obey Gauss' law. Using this formulation, the Hamiltonian per plaquette then involves only one gauge field. Employing the minimal gauge field truncation using three levels (i.e.~$d=3$), our VQE ansatz is then given by a hybrid qudit-qubit approach with one qutrit representing the gauge degree of freedom, and four qubits representing electrons and positrons residing on the four vertices, see \figrefp{fig:1}{d}. We note that these requirements for 2D connectivity and two different constituents are most easily satisfied in an all-to-all connected mixed-dimensional digital quantum processor as opposed to analog Ising-type simulators~\cite{kokail2019self}.

Our VQE ansatz is based on the Hamiltonian given in Eq.~\eqref{eq:openplaq_qubit} and reflects the underlying physics of pair creation processes: the qubit states $\ket{\downarrow} /\ket{\uparrow}$ represent vacuum/electrons on even lattice sites and positrons/vacuum on odd lattice sites, as shown in \figref{fig:encoding_si} in the \methods. The qutrit states $\{\ket{-1},\ket{0},\ket{+1}\}$ represent the electric field eigenstates of the gauge field. The circuit is initialized in the qudit-qubit state $\ket{\downarrow\uparrow\downarrow\uparrow,0}$, representing the bare vacuum $|vvvv,0\rangle$, where no particles (first four entries) or gauge field excitations (last entry) are present. 
As shown in \figref{fig:1}{d}, the two-qubit gates on the matter qubits (blue) at the beginning of the circuit populate the plaquette with electrons and positrons. When the two lattice sites directly next to the remaining gauge field are populated (see inset of \figref{fig:1}{f}), Gauss’ law requires the excitation of the gauge field to change accordingly, which is achieved by the qubit-qutrit controlled-rotation gates (red). In the final part of the circuit, four two-qubit gates (blue) adjust the matter state. This last step can be done without modifying the state of the qutrit, since the matter fields also have gauge-field-independent free dynamics. The described Hamiltonian-based VQE circuit design is extendable to larger lattices as explained in Ref.~\cite{paulson_simulating_2021}. 

Figure~\ref{fig:1}e shows a typical experimental run of the variational circuit. The resulting measured plaquette expectation values as a function of the parameter $g^{-2}$ are shown in \figrefp{fig:1}{f} for $\Omega=5$ and $m=0.1$ (see \ref{sec:methods_open_plaquette} for details on the parameter choice).The experimental data approximate the ideal result well and agree with our theoretical predictions using a simple error model, as explained in \si~\ref{sec:noise_model}. 
In the large coupling regime ($g^{-2} \ll 1$) the electric field energy term $\hat{H}_E$ dominates the Hamiltonian of Eq.~\eqref{eq:Hamiltonian}, favoring the ground state $|vvvv,0\rangle$ with $\langle\plaqop\rangle= 0$. In the weak coupling regime, on the other hand, the magnetic field energy term $\hat{H}_B$ dominates the Hamiltonian, favoring a positive vacuum plaquette expectation value ($\langle\plaqop\rangle = \frac{1}{\sqrt{2}} \approx 0.707$ for the chosen truncation). In the intermediate regime where $g^{-2} \approx 1$, there is a competition between the field-energy terms and $\hat{H}_{k}$. The ground state of the kinetic term has a positive plaquette expectation value. The presence of dynamical matter and the pair creation effect described by $\hat{H}_{k}$ leads therefore to an increase of $\langle\plaqop\rangle$ in the intermediate coupling regime, as shown in \figrefp{fig:1}{f}; this effect results in a change of the slope of the plaquette expectation value as a function of $g^{-2}$.
An alternative lattice QED model~\cite{paper1,zohar_digital_2017,raychowdhury_loop_2020,Magnifico_2021,Clemente2022,bender_variational_2023} is studied in \si~\ref{sec:model_cm}. In both cases, the relevant physics is already captured for the minimal truncation
of $d = 3$. In general, however, realizing LGT quantum computations beyond 1D will require the ability to control the gauge field dimension. We will study this key requirement in the next section.


\begin{figure*}[th!]
\centering
\includegraphics[width=0.98\textwidth]{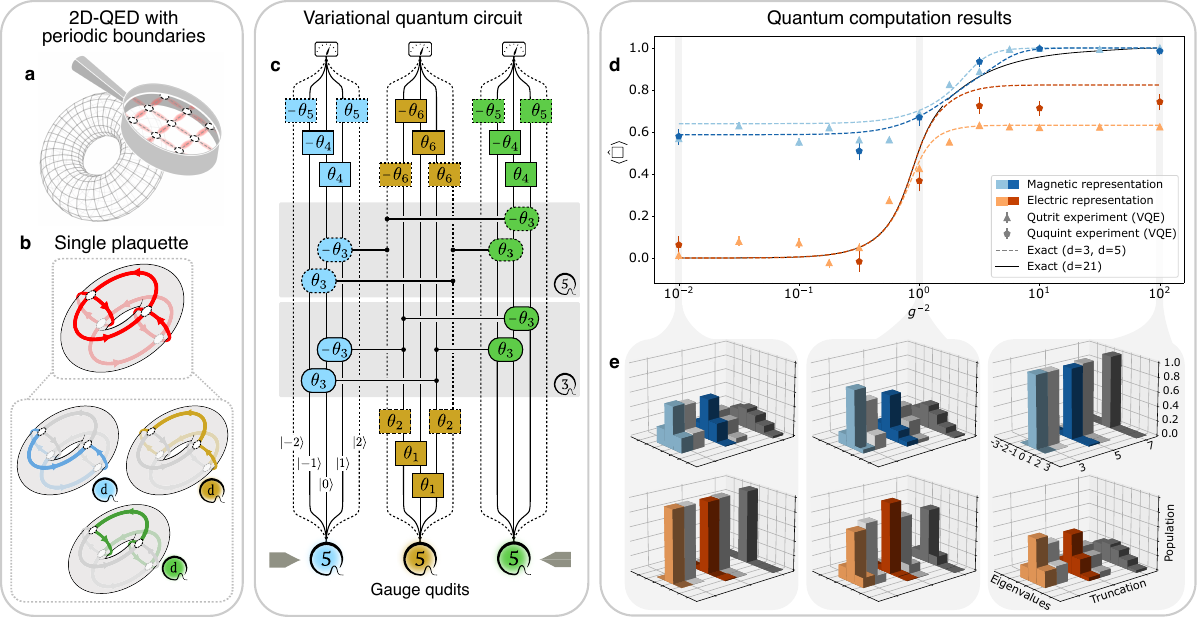}
\caption{\textbf{Refining the gauge field discretization} \textbf{a}, We consider pure gauge QED in two spatial dimensions with periodic boundary conditions, i.e.\ on a lattice on the surface of a torus. As before, the gauge field resides on the links of the lattice, while here the vertices remain empty. \textbf{b}, We consider the smallest instance of such a torus, containing four empty sites and eight gauge field links. The ground state of this particular system can be described via three separate circulation-paths of the gauge field, called rotators as discussed in \methods~\ref{sec:periodic_plaquette}. Each rotator fulfills an eigenvalue equation equivalent to a single link gauge field and can thus be subject to the same truncation rules as discussed in the main text by employing a $d$-level qudit. Here, we demonstrate the difference between a realization employing qutrits and ququints. \textbf{c}, The variational circuit in the electric representation (see main text) for the qutrit (solid lines) and the ququint (all, except shaded box marked with qutrit symbol) truncation. The explicit form of the gates employed is given in \methods~\ref{sec:circuit_gates}. \textbf{d}, Experimentally measured expectation values of the plaquette operator $\plaqop$ in the VQE-optimized ground states using qutrits (light blue and orange triangles), compared to ququints (dark blue and red pentagons). The error bars indicate one sigma statistical uncertainty. The black line represents numerical results obtained for $d=21$, using the electric (magnetic) representation for small (large) values of $g^{-2}$, while the dashed lines are exact numerical results for qutrits and ququints. \textbf{e}, The duality between the electric (all orange bars) and magnetic (all blue bars) representations is clearly seen in the experimentally measured populations of the eigenvectors of the yellow rotator from panel b for the qutrit VQE experiment and ququint experiment. The grey bars are results obtained via exact diagonalization. In the regime dominated by the electric Hamiltonian (small $g^{-2}$) a qutrit (light orange) representation is enough to approximate the correct ground state, while for larger $g^{-2}$ truncation errors become more relevant and a ququint representation (dark orange) becomes advantageous. A complementary argument applies to the magnetic qutrit (light blue) and ququint (dark blue) representation.}
\label{fig:2}
\end{figure*}


\section{Towards refining the gauge field discretization}
\label{sec:periodic}
The hardware-efficient representation of gauge fields allows us to experimentally address a second prerequisite on the path towards simulating Nature through LGT quantum computations: controlling the number of gauge field levels. More specifically, we demonstrate how our qudit platform allows us to seamlessly improve the gauge field discretization from qutrits ($d=3$) to ququints ($d=5$). As a concrete example, we study the dependence of the plaquette expectation value on the bare coupling at different discretizations, as a first important step towards quantum computations of the running of the coupling. To this end, we consider QED on a 2D lattice with periodic boundary conditions, which takes the form of a torus, see \figref{fig:2}{a}.

For our proof-of-concept demonstration, we consider a pure gauge theory $\hat{H} = g^2\hat{H}_E +(1/g^2) \hat{H}_B$, see Eq.~\eqref{eq:Hamiltonian_terms}. \figrefp{fig:2}{b} shows the minimal system consisting of four vertices and eight gauge fields (one per link). Using Gauss’ law we reduce the number of independent gauge fields to five, and it can be shown that three of these are sufficient to describe ground state properties~\cite{paper1}. Instead of describing the gauge degrees of freedom in terms of the fields associated with the individual links of the lattice as before, we switch to a more convenient description in terms of operators called ``rotators''. As explained in Ref.~\cite{paper1}, each of the three rotators in our simulation can be visualized as loops around a different plaquette, see \figrefp{fig:2}{b}.

So far, we truncated the gauge field directly in the electric field eigenbasis, i.e.\ in our first experiment we included eigenstates $|E\rangle$ of the electric field operator $\hat E$ with $\hat{E}|E\rangle = E|E\rangle$ and $E=0,\pm 1$. However, to determine the plaquette expectation value resource-efficiently across all values of the coupling, we now employ a more suitable truncation scheme that we introduced in Ref.~\cite{paper1}. Our method is based on a Fourier transformation: for large couplings (where $g^{-2}\ll1$), the Hamiltonian is dominated by the electric field contribution $\hat{H}_E$, and a gauge field truncation in the electric \mbox{(E-)} field basis is suitable, which we refer to as the electric representation. For small couplings (where $g^{-2}\gg1$), the magnetic field term dominates, and accordingly, a magnetic \mbox{(B-)} field basis (using B-field eigenstates) is more efficient. The VQE circuits for the \mbox{E-} and \mbox{B-} representations are shown in the \si~\ref{sec:periodic_circuit} and in \figrefp{fig:2}{c} respectively. As explained in more detail in the \si, their construction is inspired by the form of the Hamiltonian.

Figure~\ref{fig:2}d shows the resulting ground-state plaquette expectation values versus $g^{-2}$, along with our theoretical predictions that include a simple noise model, as described in \si~\ref{sec:noise_model}. For qutrits and ququints, we perform the full VQE as in \figref{fig:1}. Note that we show the results for both representations across all values of the coupling $g^{-2}$, even though the validity of the electric (magnetic) representation is restricted to the large (small) coupling regime, where $g^{-2}\ll1$ ($g^{-2}\gg1$). The gap between the curves in the intermediate region $g^{-2} \approx 1$, where the electric and magnetic representations perform equally, stems from the truncation of the gauge fields, see Ref.~\cite{paper1}. As the truncation is increased, the two curves rapidly approach each other and eventually agree for some intermediate value of $g^{-2}$, as indicated by the experimental data and confirmed by a numerical simulation at $d=21$, see \figrefp{fig:2}{d}. The value of $g^{-2}$ where the curves are closest indicates the point at which the representation should be switched.

This effect can also be observed in \figrefp{fig:2}{e}, where we depict the measured populations of the gauge fields in the ground state. In the large coupling regime (where $g^{-2}\ll1$), the distribution of electric field states is narrow, which allows for an efficient truncation in the E-basis. By contrast, this distribution becomes very broad in the small coupling regime, where it tends to an equally weighted superposition of infinitely many E-field levels in the limit $g^{-2}\gg1$. As a result, the accurate approximation of the ground state in the E-basis implies exploding resource costs without a basis change~\cite{paper1}. The behavior of the B-representation is complementary. The aforementioned closing of the gap between the plaquette expectation values in the E-representation (red) and the B-representation (blue) in \figrefp{fig:2}{d} thus corresponds to a better representation of the ground state in \figrefp{fig:2}{e}. The closing gap for intermediate values of $g^{-2}$ provides an indication of how well LGT quantum computations with finite $d$ approximate the untruncated results~\cite{paper1}. 

Similarly, the so-called ``freezing” of the truncated plaquette expectation value in the weak coupling regime $g^{-2}\gg 1$ serves as an indicator for how well the ground state is approximated. Freezing occurs in both classical and quantum computations if the number of levels $d$ is too small to accurately reflect the spread of the ground state wave function: since there are too few ``bins'' for the population of the gauge field, the truncated state cannot capture the fact that the ground state still changes with increasing $g^{-2}$, which is visible as a premature flattening of the plaquette expectation value vs $g$ (the dashed blue lines for $d = 3, 5$ in Fig. 2d flatten out earlier than the solid black line for $d = 21$).
A detailed explanation and analysis are provided in Ref.~\cite{paper1}, showing fast convergence of the plaquette expectation value to the true value already for $d\leq 10$. In general, different problems will permit different degrees of truncation. Yet, the moderate values of $d$ available in typical atomic systems are expected to be sufficient for addressing a range of interesting problems, particularly those involving local observables and the ground state sector of the theory~\cite{dimeglio2023quantum,beck2023quantum}.
Notably, in our experimental setup improving the gauge field discretization is achieved with only minor modifications, involving the same number of ions and entangling gates, see \si~\ref{sec:periodic_circuit}.

\section{Towards real-time dynamics}

\begin{figure}
\centering
\includegraphics[width=0.99\columnwidth]{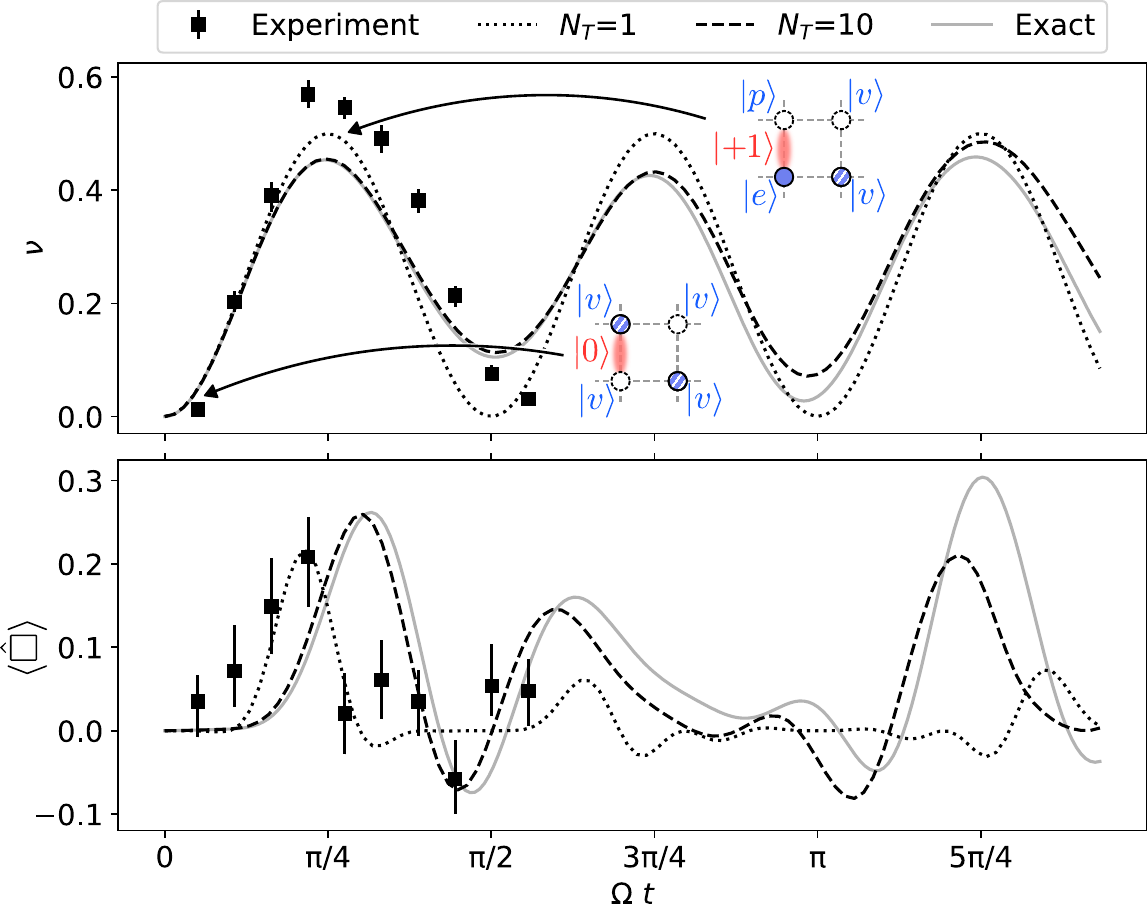}
\caption{\textbf{Time evolution for 2D-QED with matter.} Expectation values of the particle number density $\nu$ and plaquette operator $\langle\hat{\square}\rangle$ for $\Omega = 5$, $m = 0.1$, and $g^{-2} = 10^{-1.4}$. We show the numerical values for $N_T=1$ (dotted) and $N_T = 10$ (dashed) Trotter steps, with the solid line representing the exact evolution. 
	Experimental data (squares) is presented with error bars indicating one sigma statistical uncertainty. The data is obtained by post-selection on the zero magnetization sub-sector on the matter sites. Insets show examples of the dominant states at different times of the evolution.}
\label{fig:long_trotter}
\end{figure}

We complete our study of qudit LGT quantum computations by investigating the prospect of simulating real-time evolutions. We take a first step in this direction by using mixed-dimensional entangling gates to realize a digital quantum simulation (in the form of a Trotter protocol~\cite{Lloyd1996}) for the model used in Sec.~\ref{sec:open}, i.e.\ a plaquette with open boundary conditions including both gauge fields and dynamical matter.

As in the case of the VQE demonstration in Fig.~\ref{fig:1}, we study this model with a hybrid qubit-qutrit system. Starting from the bare vacuum $|vvvv, 0>$ as the initial state, we study the dynamics of the system under the Hamiltonian given in Eq.~\eqref{eq:openplaq_qubit} using a single Trotter step of varying length, see \methods~\ref{sec:realtime} for details. This time evolution can be interpreted as a quench from the strong coupling regime ($g^{-2} \ll 1$), where the bare vacuum is the ground state, to an intermediate coupling value. In the latter regime, the kinetic term of the Hamiltonian drives the creation and subsequent annihilation of particle-antiparticle pairs, leading to an increase of the mean particle number density $\nu$, followed by oscillatory behavior. Due to Gauss’ law, the creation of charged particles also affects the electromagnetic fields present in the system, which is observed as corresponding dynamics of the plaquette expectation value in Fig.~\ref{fig:long_trotter}b.

\medskip
\medskip
\section{Outlook}
\label{sec:outlook}

Qudits provide a hardware-efficient approach to quantum-simulating gauge theories. Using a universal trapped-ion qudit quantum computer with all-to-all connectivity enables us to simulate arbitrary geometries, and thus perform quantum simulations of 2D-LGTs. In contrast to 1D models, gauge fields must be included explicitly, and in contrast to condensed matter models, the gauge fields in particle physics are described by more complicated gauge groups and must have more than two states. In particular, we are able to simulate a basic building block of 2D-QED with both dynamical matter and gauge fields (Fig.~\ref{fig:1}), and to study different gauge fields discretizations (Fig.~\ref{fig:2}). These complex computations are rendered possible by high-fidelity qudit control, combined with VQE circuits that are much shallower than comparable qubit-based implementations of gauge theory calculations. While we exploit native qudit circuits to study equilibrium properties, the techniques we developed are directly adaptable to digital quantum simulations of real-time dynamics (Fig.~\ref{fig:long_trotter}).

The ultimate goal are LGT quantum simulations in three spatial dimensions. Importantly, there is a dramatic change in the simulation requirements from 1D to 2D, while there is little change going from 2D to the full 3D model~\cite{Kan2021Investigating, zohar_quantum_2021}. In particular, while the high-dimensional gauge fields can be integrated out in 1D models, they are dynamic degrees of freedom in 2D and 3D and must be simulated explicitly. 
Our results on QED simulations beyond 1D thus represent a major step towards simulating 3D LGTs. In particular, our protocol based on eliminating redundant gauge degrees of freedom employed here can be directly extended to 3D (in the case of QED shown in Ref.~\cite{Kan2021Investigating}). Our qudit techniques can be applied to virtually all other quantum computing architectures and hardware platforms. For all of them, the remaining major task is scaling up the system sizes, which makes an efficient gauge field representation even more important. Beyond just programmable local dimensions, qudit-based systems enjoy larger freedom for designing interactions to optimally match the target problem. Notably, quantum error correction schemes, which are making great progress in conventional qubit systems~\cite{bluvstein_logical_2024, da_silva_demonstration_2024, campbell_series_2024} are compatible~\cite{watson_fast_2015} with our hardware-efficient qudit approach. Our demonstration of a qudit-based quantum simulation of high-energy physics phenomena thus paves the way to a new generation of qudit-based applications in all areas of quantum technology.

\section*{Acknowledgements}

We thank Karl Jansen and Arianna Crippa for helpful discussions on various lattice QED models. This research was funded by the European Union under the Horizon Europe Programme---Grant Agreements 101080086---NeQST and 101113690---PASQuanS2.1, by the European Research Council (ERC, QUDITS, 101080086), and by the European Union’s Horizon Europe research and innovation programme under grant agreement No 101114305 (``MILLENION-SGA1'' EU Project). Views and opinions expressed are however those of the author(s) only and do not necessarily reflect those of the European Union or the European Research Council Executive Agency. Neither the European Union nor the granting authority can be held responsible for them. We also acknowledge support by the Austrian Science Fund (FWF) through the SFB BeyondC (FWF Project No. F7109) and the EU-QUANTERA project TNiSQ (N-6001), by the Austrian
Research Promotion Agency (FFG) through contracts 897481 and 877616, and by the IQI GmbH.
We further received support by the ERC Synergy
Grant HyperQ (grant number 856432), the BMBF
project SPINNING (FKZ:13N16215) and the EPSRC grant EP/W028301/1.
This research was also supported by the Natural Sciences and Engineering Research
Council of Canada (NSERC), the Canada First Research Excellence Fund (CFREF, Transformative Quantum Technologies), New Frontiers in Research Fund (NFRF),
Ontario Early Researcher Award, and the Canadian Institute for Advanced Research (CIFAR).\\
\noindent\textbf{Author contributions} JZ, JFH, AJJ, LD, PZ, and CM developed the theory results. MM, CE, LP, AS, RB, TM, PS, and MR performed the experiments. MM and JZ analyzed the data and performed numerical simulations. CM and MR supervised the project. All authors contributed to writing the manuscript. \\
\noindent\textbf{Competing Interests} The authors declare no competing interests.\\
\noindent\textbf{Data availability} The data underlying this work are available at \url{https://doi.org/}.\\
\noindent\textbf{Code availability}
Code used for data analysis is available from the corresponding author upon reasonable request.

\newpage
\bibliography{bibliography.bib}

\newpage
\clearpage
\appendix


\section{Realizing Controlled Rotations in Qudits}
\label{sec:experimetal_details}

\begin{figure}[ht]
	\centering
	\includegraphics[scale=1.0]{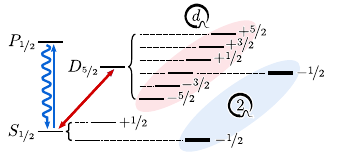}
	\caption{Simplified level scheme of $^{40}\mathrm{Ca}^+$. Doppler cooling and state detection is performed on the short-lived $S_{\nicefrac{1}{2}}\leftrightarrow P_{\nicefrac{1}{2}}$ transition with a life time of $\approx 9~\mathrm{ns}$. \textit{Qubits} are encoded in the Zeeman states $\ket{0} = S_{m = -\nicefrac{1}{2}}$ and $\ket{1} = D_{m = -\nicefrac{1}{2}}$, while \textit{qudits} are encoded only in the $D_{\nicefrac{5}{2}}$ manifold. Any excitation of a $D_{\nicefrac{5}{2}}$ state decays in $T_1\approx 1.1~\mathrm{s}$ to the $S_{\nicefrac{1}{2}}$ ground states; this transition is addressed by a narrow-band laser with a coherence time of $T_2 = 92(9)~\mathrm{ms}$.}
	\label{fig:exp_ca40_scheme}
\end{figure}

We encode each qudit in Zeeman states of trapped $^{40}\mathrm{Ca}^+$ ions and manipulate their quantum state by sequences of laser pulses. Doppler cooling and state detection is performed by driving the short-lived $S_{\nicefrac{1}{2}}\longleftrightarrow P_{\nicefrac{1}{2}}$ transition, monitoring the fluorescence of the individual ions on a CCD camera; a simplified level scheme of the relevant states is shown in Fig.~\ref{fig:exp_ca40_scheme}. Quantum gate operations are implemented by selectively addressing single -- or arbitrary pairs of -- ions via a high numerical aperture objective, coupling the $D_ {\nicefrac{5}{2}}$ manifold with the $S_{\nicefrac{1}{2}}$ ground states \cite{AddressingKim_2008}. Due to the geometry of our ion trap, the beam is aligned at an angle of $22.5^\circ$ with respect to the trap axis, leading to a Lamb-Dicke factor of $\eta = 0.041$ for an axial trap frequency of $\omega_z = 0.77~\mathrm{MHz}$. We measure a heating rate of $2.7(2)$ phonons per second and a motional coherence time $\tau$ of $27.4(4)~\mathrm{ms}$.

Qudits with dimension $d = 2$ (i.e.\ \textit{qubits}) are encoded, such that $\ket{0} = S_{\nicefrac{1}{2}}(m = -1/2)$ and $\ket{1} = D_{\nicefrac{5}{2}}(m = -1/2)$ as shown in Fig.~\ref{fig:exp_ca40_scheme}. For $d > 2$ only the $D_{\nicefrac{5}{2}}$ manifold of $^{40}\mathrm{Ca}^+$ is used for encoding the qudit, while the $S_{1/2}$ ground states act as auxilliary levels $\ket{g}$, which are only populated during gate operations and readout.

Mixed-dimensional qudit-qudit interactions are engineered by controllably coupling qudits to a single phonon mode, effectively, using an Anti-Jaynes-Cummings Hamiltonian described by
\begin{equation}
	\hat{H}_j = i\eta\Omega_j\left(\hat{a}^\dagger\hat{\sigma}^{+}_j + \hat{a}\hat{\sigma}^{-}_j\right)^{\ket{g}\longleftrightarrow\ket{k}},
	\label{eq:AJC_hamiltonian}
\end{equation}
on the $j$-th ion with Rabi frequency $\Omega_j$. Here, $\hat{a}^\dagger\hat{\sigma}^+_j$ excites the qudit from the (auxilliary) ground state $\ket{g}$ from the $S_{\nicefrac{1}{2}}$ manifold to a state $\ket{k}$ in the $D_{\nicefrac{5}{2}}$ manifold, simultaneously injecting a phonon into the motional mode; $\hat{a}\hat{\sigma}^{-}_j$ has the opposite effect. In our setup, we realize this interaction by driving controlled laser pulses $B(\theta, \phi)$ tuned to the first (\textit{blue}) sideband (BSB) of the axial center-of-mass (COM) motional mode. This mode is favorable due to the homogenous coupling for all ions in the string and the reduced calibration effort, as all laser pulses have close to identical parameters for all qudits. 

\begin{figure}[ht]
	\centering
	\includegraphics[scale=1.05]{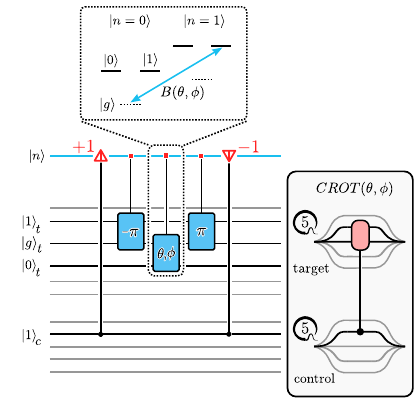}
	\caption{Qudit circuit for implementing a mixed-dimensional controlled rotation (C-ROT) gate. For each 5-dimensional qudit, we consider a two-level subspace, containing the states $\ket{0}$ and $\ket{1}$, coupled to an auxiliary ground state $\ket{g}$. The conditional interaction with the phonon mode (blue) depending on the control state is shown by the red triangles, with its orientation indicating the creation / annihilation of a phonon. If the motional mode is excited, three BSB pulses act locally on the target qudit, realizing the rotation. At the end of the sequence, the qudits are again disentangled from the motion. In the inset on the right, we introduce a symbol for the C-ROT operation $CROT(\theta, \phi)$, which allows us to draw controlled qudit operations in quantum circuits in a similar manner to qubit gates, see e.g.\ \figrefp{fig:1}{d}.}    
	\label{fig:exp_crot_sequence}
\end{figure}

Envisioned to realize the Cirac-Zoller controlled-NOT (CNOT) quantum gate in trapped ions \cite{Cirac1995}, sequences of local $B(\theta,\phi)$ pulses are well suited to implement mixed-dimensional controlled rotations (C-ROTs) in qudits as shown in Fig.~\ref{fig:exp_crot_sequence}. Here, we consider a two-level subspace of a qudit, described by the states $\ket{0}$ and $\ket{1}$, which can be coupled to the motional mode $\ket{n}$ via the auxiliary ground state $\ket{g}$. Conditioned on the control state -- here $\ket{1}$ --  the \emph{control} qudit is entangled with the motion by the interaction in Eq.~\eqref{eq:AJC_hamiltonian}, creating a phonon as indicated by the upward triangle in Fig.~\ref{fig:exp_crot_sequence}. To test the control state, a resonant $\pi$-pulse transfers the population in $\ket{1}$ to $\ket{g}$, before a second $\pi$-pulse on the blue sideband is applied. It is evident that this operation has no effect if the \emph{control qudit} is initially not in the state $\ket{1}$. On the \emph{target} qudit, we apply a sequence of three sideband pulses (blue boxes), namely
\begin{equation*}
	B(\pi,0)^{\ket{g}\to\ket{1}} B(\theta, \phi)^{\ket{g}\leftrightarrow\ket{0}} B(-\pi, 0)^{\ket{1}\to\ket{g}},
\end{equation*}
where the superscripts indicate the respective coupling between the state $\ket{g}$ and the states $\ket{0},\ket{1}$. Crucially, these operations affect the \emph{target} qudit only if the phonon mode is excited. In the final step, the operations on the \emph{control} qudit are applied in reverse order, disentangling the \emph{control} qudit from the COM mode, annihilating a phonon depicted by the downward facing triangle. 

We extend the well-established circuit representation of quantum gates in qubits by introducing a way to draw interactions in qudits in a similar manner. As shown in the inset in Fig.~\ref{fig:exp_crot_sequence}, we expand the line representing each qudit into $d$ individual rails, allowing us to draw gates acting on subspaces of the qudit unambiguously, and contract the rails back to a single line after the subspace gate. In the case of C-ROTs, the control state is indicated in the usual fashion.

\begin{figure}[htb]
	\centering
	\includegraphics[scale=1.05]{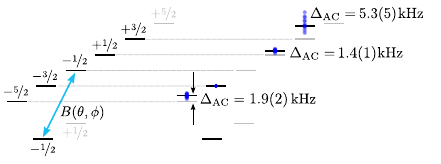}
	\caption{AC stark shifts on spectator levels for an excitation of the blue sideband of the $S_{-1/2}\leftrightarrow D_{-1/2}$ transition. While the AC Stark shift on this transition is directly compensated by a second, off-resonant laser beam, the spectator levels $D_{-5/2}$, $D_{+1/2}$ and $D_{+3/2}$ remain shifted by a few $\mathrm{kHz}$ with respect to the $S_{-1/2}$ ground state; we measure no shift for the $D_{-3/2}$ state. The individual blue dots correspond to different runs over the course of several days and the uncertainty represents one standard deviation of the fit uncertainty. These spectator shifts are compensated in software by storing a phase register for each qudit state and phase-shifting the subsequent operations on the affected qudit transitions by an appropriate amount.}
	\label{fig:exp_stark_shifts_spectators}
\end{figure}

As the Rabi frequency on the BSB is proportional to the Lamb-Dicke factor $\eta$, driving the blue sideband requires substantially more laser power than carrier transitions. Each $B(\theta, \phi)$ will thus introduce unwanted AC Stark shifts $\Delta_{\mathrm{AC}}$ on the order of a few $\mathrm{kHz}$~\cite{Ringbauer2022A-universal}, which must be carefully taken into account for achieving high-fidelity C-ROT gates. We compensate these shifts using a two-fold approach: (1) shifts on the actively driven BSB transition are compensated by an off-resonant second-beam technique~\cite{Haeffner_2003, Meth2022}, while (2) additional shifts on spectator \textit{carrier} transitions $\ket{l \neq k}$ are measured and compensated in software by frame updates on subsequent operations. An example of shifted spectator states is shown in Fig.~\ref{fig:exp_stark_shifts_spectators} for $B(\theta, \phi)$ on the blue sideband of the transition $S_{-1/2}\leftrightarrow D_{-1/2}$ with a Rabi frequency $2\pi\cdot 4~\mathrm{kHz}$. For each state $\Delta_{\mathrm{AC}}$ is obtained by a Ramsey measurement. 

\section{2D-QED Hamiltonian}
\label{sec:Methods_Hamiltonian}
The models we simulate are instances of lattice QED and are defined on a two-dimensional discretization of space. Matter (electrons and positrons) and gauge bosons (photons) are defined on the sites and on the links of this lattice respectively, as shown in \figrefp{fig:1}{a}.

\begin{figure}[htb]
	\centering
	\includegraphics[width=0.95\columnwidth]{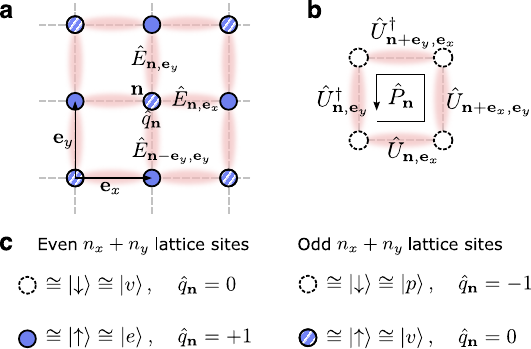}
	\caption{\textbf{Elements of two-dimensional lattice QED.} \textbf{a}, 2D lattice for the Kogut-Susskind Hamiltonian formulation. On each site $\vec{n}=(n_x, n_y)$ resides a fermionic field operator $\hat{\phi}_{\vec{n}}$ that represents 
		the presence or absence of particles. Each site can either be in the vacuum state or occupied by an electron (positron) for even (odd) sites. Electrons (positrons) are shown as filled circles with solid (striped) color, see panel c. The gauge bosons are described by operators on the links of the lattice. \textbf{b}, The operator $\hat{P}_{\vec{n}}$ is defined around a counterclockwise path around the plaquette with origin in $\vec{n}$. For every link, $\hat{P}_{\vec{n}}$ acquires a factor $\hat{U}$ if the path goes towards the positive directions $\vec{e}_{\mu}$, with $\mu \in \{x,y\}$, and a factor $\hat{U}^{\dagger}$ otherwise. \textbf{c}, Mapping between fermionic states, qubit states (obtained via a Jordan-Wigner transformation), and physical states, including associated charges.}
	\label{fig:encoding_si}
\end{figure}
Here, matter is described by fermionic field operators 
$\hat{\phi}_{\vec{n}}$ with spatial label $\vec{n} = (n_x, n_y)$. 
We employ the staggered formulation~\cite{kogut1975hamiltonian}, in which lattice sites can either be in the vacuum state or host electrons (positrons) residing on even (odd) lattice sites, carrying a $+1$ ($-1$) charge. The charge in terms of the fermionic field is given by $\hat{q}_{\vec{n}} = \hat{\phi}_{\vec{n}}^{\dagger}\hat{\phi}_{\vec{n}} - \frac{1-(-1)^{n_x + n_y}}{2}$.

The gauge field residing on the links between each pair of sites is described by an electric field operator $ \hat{E}_{\vec{n},\vec{e}_\mu}$ ($\mu \in \{x,y\}$) that possesses an infinite but discrete spectrum $\hat{E}_{\vec{n},\vec{e}_\mu} \ket{E_{\vec{n},\vec{e}_\mu}} = E_{\vec{n},\vec{e}_\mu} \ket{E_{\vec{n},\vec{e}_\mu}}$, where $E_{\vec{n},\vec{e}_\mu} = 0,\, \pm1,\,\pm2,\dots$. The link operator $\hat{U}_{\vec{n},\vec{e}_\mu}$ acts as a lowering operator for the electric field: 
\begin{equation}
	\label{eq:link_operator}
	\hat{U}_{\vec{n},\vec{e}_\mu} \ket{E_{\vec{n},\vec{e}_\mu}} = \ket{E_{\vec{n},\vec{e}_\mu} - 1}.
\end{equation}
The total Hamiltonian
	\begin{eqnarray}\nonumber
		\hat{H} = g^2\hat{H}_E +\frac{1}{g^2} \hat{H}_B + m\hat{H}_m + \Omega\hat{H}_k 
	\end{eqnarray}

\noindent is then given by a sum of four parts as shown in Eq.~\eqref{eq:Hamiltonian}. Note that compared to the convention used in~\cite{paper1,paulson_simulating_2021} we have opted to redefine the four Hamiltonians such that they are independent of the mass and coupling constant. Moreover, we consider here another formulation of lattice QED~\cite{wiese2013ultracold, crippa_jansen_2024}, in which the signs in the kinetic term are different from the ones used in ~\cite{paper1,paulson_simulating_2021}. For a discussion of the experimental results for the model employed in~\cite{paper1,paulson_simulating_2021}  see App.~\ref{sec:model_cm}. By employing the Kogut-Susskind formulation in natural units $\hbar=c=1$, and with lattice spacing $a=1$, the terms in the Hamiltonian read~\cite{crippa_towards_2024}
\begin{subequations}\label{eq:Hamiltonian_terms}
	\begin{align}
		\hat{H}_E &= \frac{1}{2}\sum_{\vec{n}}\left(\hat{E}_{\vec{n},\vec{e}_x}^2 + \hat{E}_{\vec{n},\vec{e}_y}^2 \right), \\
		\hat{H}_B &= -\frac{1}{2}\sum_{\vec{n}}\left(\hat{P}_{\vec{n}} + \hat{P}_{\vec{n}}^\dagger \right), \\
		\hat{H}_m &= \sum_{\vec{n}} (-1)^{n_x+n_y} \hat{\phi}_\vec{n}^\dagger \hat{\phi}_{\vec{n}}, \\
		\begin{split}
			\hat{H}_k &=  \sum_{\vec{n}} \bigg( i \hat{\phi}_{\vec{n}}^{\dagger} \hat{U}_{\vec{n},\vec{e}_x}^\dagger \hat{\phi}_{\vec{n}+\vec{e}_x} + \\
			& \quad (-1)^{n_x + n_y + 1} \hat{\phi}_\vec{n}^\dagger \hat{U}_{\vec{n},\vec{e}_y}^\dagger \hat{\phi}_{\vec{n}+\vec{e}_y}  + \mathrm{H.c.} \bigg).
		\end{split}
	\end{align}
\end{subequations}

\noindent
$\hat{H}_E$ and $\hat{H}_B$ represent the free electric and magnetic field, while $\hat{H}_m$ and $\hat{H}_k$ describe the free matter field and its interaction with the gauge field. The operator $\hat{P}_{\vec{n}}$ is defined on a counter-clockwise closed loop around the plaquette with origin $\vec{n}$ (see \figrefp{fig:encoding_si}{b}) as
\begin{equation}
	\hat{P}_{\vec{n}} = \hat{U}_{\vec{n},\vec{e}_x}\hat{U}_{\vec{n}+\vec{e}_x,\vec{e}_y}\hat{U}_{\vec{n}+\vec{e}_y,\vec{e}_x}^\dagger\hat{U}_{\vec{n},\vec{e}_y}^\dagger.
	\label{eq:P_op}
\end{equation}
Noting that the link operator $\hat{U}_{\vec{n},\vec{e}_\mu}$ is expressed in terms of the gauge fields' vector potential $\hat{A}_{\vec{n},\vec{e}_\mu}$, i.e.\ $\hat{U}_{\vec{n}, \vec{e}_\mu}\sim \mathrm{exp}\lbrace i g\hat{A}_{\vec{n},\vec{e}_\mu} \rbrace$, one observes that the exponent of $\hat{P}_\vec{n}$ forms a discrete lattice curl of the vector potential. The plaquette operator $\plaqop = \frac{1}{2V}\sum_{\vec{n}}\left(\hat{P}_{\vec{n}} + \hat{P}_{\vec{n}}^\dagger \right)$, where $V$ is the number of plaquettes, is therefore proportional to the magnetic field energy and a true multi-dimensional quantity which has no analog in 1D-QED. In quantum field theories the spontaneous creation of particle/antiparticle pairs in vacuum means that, when measuring the strength of a charge, the result can depend on the distance and energy scale at which it is probed. Since the coupling is proportional to the charge~\cite{hamer1997series}, this means that the physical coupling, which is an important parameter in phenomenological HEP models, also depends on the energy scale. This phenomenon is known as the running of the coupling. The dependence of the ground-state plaquette expectation value $\langle \plaqop\rangle$ on $g^{-2}$, where $g$ is the bare coupling, can be related to the running of the coupling and it is discussed more in detail in~\cite{Clemente2022}.

Importantly, not all quantum states in the considered Hilbert space are physical. The gauge field and charge configurations of physical states $|\Psi_{\text{phys}}\rangle$ have to fulfill Gauss' law at every site. Here, the familiar law from classical electrodynamics
$\nabla \vec{E}(\vec{r})-\rho(\vec{r})=0$, where $\rho(\vec{r})$ is the charge density at point $\vec{r}$, takes the form 
\begin{equation}
	\hat{G}_{\vec{n}}|\Psi_{\text{phys}}\rangle=0,
	\label{eq:gausslaw}
\end{equation}
where the Gauss' operator is defined as
\begin{equation}
	\hat{G}_{\vec{n}} = \sum_\mu (\hat{E}_{\vec{n},\vec{e}_\mu} - \hat{E}_{\vec{n}-\vec{e}_\mu,\vec{e}_\mu}) -  \hat{q}_\vec{n}.
	\label{eq:gauss_operator}
\end{equation}
In general, Gauss' law requires physical states to be eigenvectors of the Gauss operator; we have made the choice in Eq.~\eqref{eq:gausslaw} to consider the eigenvalues to be zero, which describes a model with no external charges. The total charge $\sum_\vec{n} \hat{q}_\vec{n}$ can be shown to also be a symmetry of the Hamiltonian. In this work, we choose to study states that have zero total charge.

\section{Encoded Hamiltonian}\label{sec:Methods_EncodedHamiltonian}
In the following, we provide the Hamiltonians studied via the VQE circuits given in \figrefp{fig:1}{d} and \figrefp{fig:2}{c}. 
\newline

\subsection{2D-QED with Matter}
\label{sec:methods_open_plaquette}
Let us consider a single plaquette with open boundary conditions with origin in $\vec{n} = (0, 0)$ and extending in the positive directions, as shown in the inset of \figrefp{fig:1}{f}. As was discussed in \cite{paulson_simulating_2021}, Gauss' law given in Eq.~\eqref{eq:gausslaw} can be used to eliminate three gauge fields, resulting in an Hamiltonian that contains four fermionic fields and one gauge field, which we choose to be the one between sites $(0, 0)$ and $(0, 1)$. We encode the fermions i$, \dots, $ iv (conventionally ordered clockwise around the plaquette starting from $(0, 0)$) into a chain of qubits indexed by $1, \dots, 4$ respectively by applying the following Jordan-Wigner transformation 
\begin{equation}
	\label{eq:JW_transformation}
	\hat{\phi}_{i} = \prod_{j<i} (e^{i\alpha_j}\hat{\sigma}_{j}^{z}) \hat{\sigma}_{i}^{-}, \quad \hat{\phi}_{i}^{\dagger} = \prod_{j<i} (e^{-i\alpha_j}\hat{\sigma}_{j}^{z}) \hat{\sigma}_{i}^{+},
\end{equation}
with $\alpha_1=\alpha_3=0$, and $\alpha_2=\frac{\pi}{2}$; note that this choice of phases generates a qubit Hamiltonian with all real coefficients.

In particular, the charge associated to each site is now given by $\hat{q}_{i} = \frac{\hat{\sigma}_{i}^{z} + (-1)^{i+1}}{2}$. A table summarizing how the spin states are related to the fermionic states is given in~\figrefp{fig:encoding_si}{c}.
The Hamiltonian

	\begin{eqnarray}\nonumber
		\hat{H} = g^2\hat{H}_E +\frac{1}{g^2} \hat{H}_B + m\hat{H}_m + \Omega\hat{H}_k,  
	\end{eqnarray}

after the Jordan-Wigner transformation reads
\begin{subequations}
	\label{eq:openplaq_qubit}
	\begin{align}
		\begin{split}
			\hat{H}_{E} &= \frac{1}{4} ( 8 \hat{E}^{2} + 2\hat{E}
			\left( - 2 \hat{\sigma}^{z}_{1} + \hat{\sigma}^{z}_{2} - \hat{\sigma}^{z}_{4} - 2 \right) \\
			& \qquad \quad + \hat{\sigma}^{z}_{1} - \hat{\sigma}^{z}_{2} + \hat{\sigma}^{z}_{1} \hat{\sigma}^{z}_{4} + 3 ), 
		\end{split}\\
		\hat{H}_{B} &= -\frac{1}{2} \left( \hat{U} + \hat{U}^{\dagger}\right), \\
		\hat{H}_{m} &= \frac{1}{2} \left( \hat{\sigma}^{z}_{1} - \hat{\sigma}^{z}_{2} + \hat{\sigma}^{z}_{3} - \hat{\sigma}^{z}_{4} \right), \\
		\hat{H}_{k} &= \hat{\sigma}^{+}_{1} \hat{U}^\dagger \hat{\sigma}^{-}_{2} + \hat{\sigma}^{+}_{2} \hat{\sigma}^{-}_{3} - \hat{\sigma}^{+}_{4} \hat{\sigma}^{-}_{3} - \hat{\sigma}^{+}_{1} \hat{\sigma}^{-}_{4} + \text{H.c.},
	\end{align}
\end{subequations}
\noindent
where we have simplified terms in the kinetic Hamiltonian, by using the fact that we consider only states with zero total charge (and therefore zero total magnetization for the qubits), as discussed below Eq.~\eqref{eq:gausslaw}.
The gauge degree of freedom is encoded in a qutrit, and following Ref.~\cite{paulson_simulating_2021}, we truncate the gauge field operators as
\begin{equation}
	\hat{E} = \begin{pmatrix}
		1 & 0 & 0 \\
		0 & 0 & 0 \\
		0 & 0 & -1\end{pmatrix}, \qquad
	\hat{U} = \begin{pmatrix}
		0 & 0 & 0 \\
		1 & 0 & 0 \\
		0 & 1 & 0\end{pmatrix}.
\end{equation}
For the experiment in Fig.~\ref{fig:1} in the main text, we chose the parameter values $m=0.1$ and $\Omega = 5$. This choice positions our demonstration in the non-perturbative regime, where pair-creation has a significant effect. The definition of the different prefactors of the individual terms in the Hamiltonian can be found in App.~A of \cite{paper1}. 

\subsection{Pure Gauge 2D-QED}
\label{sec:periodic_plaquette}
The second simulated model is a pure gauge theory with periodic boundary conditions as shown in~\figrefp{fig:2}{a}. We consider here the minimal instance consisting of four vertices, as depicted in \figrefp{fig:2}{b}. 
As discussed in Ref.~\cite{paper1}, Gauss' law can be used to reduce the system to three independent degrees of freedom; these are described by three operators $\hat{P}_i$ ($i=1,2,3$) which are defined in Eq.~\eqref{eq:P_op}, and are each associated to the magnetic energy of a plaquette depicted in the lower part of \figrefp{fig:2}{b}. For each plaquette we define the corresponding rotator operator as the circulation of the gauge field going counterclockwise. It can be shown that the rotators have integer spectrum
\begin{equation}
	\hat{R}_i \ket{r_i} = \ket{r_i}, r_i \in \mathbb{Z},
	\label{eq:rotator_spectrum}
\end{equation}
and that the operators $\hat{P}_i$ act as lowering operators on these states, i.e.\ $\hat{P}_i\ket{r_i} = \ket{r_i - 1}$.

In order to study this model on a quantum computer, the infinite-dimensional Hilbert spaces of the rotators need to be truncated with a strategy that allows one to systematically increase the truncation size and approach the continuum limit,  as discussed in detail in Ref.~\cite{paper1}. We summarize here the essential points of the derivation.
As a first step the gauge group $\operatorname{U(1)}$ is substituted by the discrete group $\mathbb{Z}_{2L+1}$, and the Hilbert space is then truncated for each rotator to the $2l + 1$ eigenstates $\ket{-l}, \dots, \ket{l}$, for some $l \leq L$. The action of the truncated $\hat{P}_i$ operator is then given by 
\begin{equation}
	\hat{P}_i \ket{r_i} = 
	\begin{cases}
		\ket{r_i - 1}, \text{if } r_j > -l \\
		\delta_{lL} \ket{l}, \text{if } r_j = -l.
	\end{cases}
\end{equation} 
The Hamiltonian in terms of the rotators reads
\begin{subequations} \label{eq:elec_rep}
	\begin{align}
		\hat{H}^{\text{(e)}} &= g^2 \hat{H}^{\text{(e)}}_E + \frac{1}{g^2}\hat{H}^{\text{(e)}}_B, \\
		\hat{H}^{\text{(e)}}_E &= 2 \left[ \hat{R}_{1}^{2} + \hat{R}_{2}^{2} + \hat{R}_{3}^{2} - \hat{R}_{2} \left( \hat{R}_{1} + \hat{R}_{3} \right) \right],\\
		\hat{H}^{\text{(e)}}_B &= - \frac{1}{2} \left( \hat{P}_{1} + \hat{P}_{2} + \hat{P}_{3} + \hat{P}_{1} \hat{P}_{2} \hat{P}_{3} + \text{H.c.} \right),
	\end{align}
\end{subequations}
where we have used the superscript (e) to indicate that we are using here the so-called electric representation, in which the electric Hamiltonian $\hat{H}^{\text{(e)}}_E$ is diagonal. The parameter $g$ is the bare coupling and is proportional to the bare electric charge~\cite{hamer1997series}.

In the regime where $g \ll 1$, it is more convenient to apply a discrete Fourier transform to the Hamiltonian, which leads to a magnetic representation, where $\hat{H}^{\text{(b)}}_B$ is diagonal. For the calculations, let us define the following coefficients, defined in terms of the polygamma functions $\psi_{\nu}(x)$
\begin{subequations}
	\begin{align}
		f_{\nu}^{s} &= \frac{(-1)^{\nu + 1}}{2 \pi} \left[ \psi_{0}\!\left( \frac{2+ L 1 + \nu}{2(2L+1)} \right) - \psi_{0}\!\left( \frac{\nu}{2(2L+1)} \right) \right], \\
		f_{\nu}^{c} &= \frac{(-1)^{\nu}}{4 \pi^{2}} \left[  \psi_{1}\!\left( \frac{\nu}{2(2L+1)} \right) - \psi_{1}\!\left( \frac{2L+1 + \nu}{2(2L+1)} \right) \right],
	\end{align}
\end{subequations}
and introduce the notation $\ket{\bm{r}} = \ket{r_{1}}\ket{r_{2}}\ket{r_{3}}$.
The Hamiltonian in the magnetic representation is then given by
\begin{subequations}
	\begin{align}
		\hat{H}^{\text{(b)}} &= g^2 \hat{H}^{\text{(b)}}_E + \frac{1}{g^2}\hat{H}^{\text{(b)}}_B, \\
		\begin{split}
			\hat{H}^{\text{(b)}}_E &= \sum_{\nu=1}^{2L} \bigg[ f_{\nu}^{c} \left( \hat{P}_{1}^{\nu} + \hat{P}_{2}^{\nu} + \hat{P}_{3}^{\nu}  \right) \\
			& \quad +\frac{f_{\nu}}{2} \left( \hat{P}_{2}^{\nu} - \left( \hat{P}_{2}^\dagger \right)^{\nu} \right) \sum_{\mu=1}^{2L} f_{\mu}^{s} \left( \hat{P}_{1}^{\mu} + \hat{P}_{3}^{\mu} \right) \bigg] + \text{H.c.}, 
		\end{split}\\
		\begin{split}
			\hat{H}^{\text{(b)}}_B &= - \sum_{\bm{r} = - \bm{L}}^{\bm{L}} \Bigg[ \cos\!\left( \frac{2\pi r_{1}}{2L+1} \right) + \cos\! \left( \frac{2\pi r_2}{2L+1} \right) \\
			& \qquad + \cos\!\left( \frac{2\pi r_3}{2L+1}  \right) + \cos\! \left( \frac{2\pi(r_{1} + r_{2} + r_{3})}{2L+1} \right) \Bigg] \ketbra{\bm{r}}.
		\end{split}
	\end{align}
	\label{eq:magn_rep}
\end{subequations}
For our quantum calculations shown in \figref{fig:2} we choose $(L, l) = (2, 1)$ for the qutrit experiment, and $(L, l) = (3, 2)$ for the ququint experiment. Throughout the paper we use $L=l+1$.

\section{A Variational Quantum Eigensolver for Qudits}

\subsection{Variational Circuit}
\label{sec:circuit_gates}
As explained in the main text, the variational circuit shown in~\figrefp{fig:1}{d} is designed based on the form of the target Hamiltonian given in Eqs.~\eqref{eq:openplaq_qubit}. Since all the coefficients of the Hamiltonian are real, it is convenient to use gates in the variational circuit that rotate between states with real coefficients only. This approach also allows us to formulate a circuit that is more efficient in the number of variational parameters. In particular, the square blue gates between the qubits in 
\figrefp{fig:1}{d} are the magnetization-conserving gates studied in Ref.~\cite{gard_efficient_2020}, defined as 
\begin{equation}
	\hat{A}(\theta, \phi) = 
	\left(
	\begin{matrix}
		1 & 0 & 0 & 0 \\
		0 & \cos \theta & e^{i\phi} \sin \theta & 0 \\
		0 & e^{-i\phi} \sin \theta & - \cos \theta & 0 \\
		0 & 0 & 0 & 1\\
	\end{matrix}
	\right).
\end{equation}
We choose $\phi=0$ for all gates, which results in a transformation with real coefficients, and the qubits on which $\hat{A}$ acts are ordered by decreasing index.

The entangling gates in~\figrefp{fig:1}{d}, \figrefp{fig:2}{c}, \figref{fig:open_ququint}, and~\figref{fig:periodic_magnetic_circuit} marked by the rounded box are controlled rotations. When the control is active, the rotation on the target qudit is given by 
\begin{equation}
	\hat{R}^{i,j}_{y}(\theta) = \exp \left( - i \theta \hat{\sigma}^{i,j}_{y} /2 \right),
\end{equation}
where $i,j$ are the levels addressed in the target qudit, and $\hat{\sigma}^{i,j}_{y}$ is the corresponding Pauli $y$ matrix; this choice ensures that the rotation has real coefficients. In the qubit-qudit configuration, the rotation is active when the control qubit is in the $\ket{\downarrow}$ state, while for the qudit-qudit gate, the control state is explicitly marked in the corresponding circuits. The experimental implementation of these gates is discussed in more detail in Sec.~\ref{sec:experimetal_details}. 
The last type of gate used in our variational circuit is the qudit-internal rotation shown in \figrefp{fig:2}{c} and~\figref{fig:periodic_magnetic_circuit} as square boxes. They are given by the $x$ rotation between the addressed states $i,j$
\begin{equation}
	\hat{R}^{i,j}_{x}(\theta) = \exp \left( - i \theta \hat{\sigma}^{i,j}_{x} \right).
\end{equation}

\subsection{Optimization Algorithm}
\label{sec:vqe}
For our VQE experiments, we employ two optimization strategies, which are both based on Bayesian optimization (BO)\cite{Frazier2018aa}. Here, the algorithm collects evaluations of the cost function
$\langle\hat{H}(\boldsymbol{\theta})\rangle$ for proposed sets of $n$ input parameters $\boldsymbol{\theta} = \lbrace\theta_i\rbrace_{i=1\dotsc n}$ and constructs a surrogate model via a Gaussian process to represent the cost function. This allows one to quantify how likely further evaluations of the cost function are to achieve an improvement over the best-found value so far and hence reduces the overall number of evaluations to be performed. Importantly, the method does not require gradient evaluations and can tolerate noisy input data.

For the system with periodic boundary conditions in Fig.~2 of the main text, our BO strategy is very close to the one described in Ref.~\cite{Frazier2018aa}. We initialize the optimizer by evaluating $\langle\hat{H}\rangle$ at the corners of the parameter space, creating a $n\times n$ grid. Next, the algorithm maps out the energy landscape by tuning $\boldsymbol{\theta}$ for up to 100 evaluations (in the qutrit case). Treating the input parameters as vectors $\vec{v}(\boldsymbol{\theta})$, convergence is achieved if the norm $\abs{\vec{v}(\boldsymbol{\theta}) - \vec{v}(\boldsymbol{\theta}')}$ between 5 consecutive runs does not exceed a threshold of $0.01$.

The variational optimization for the case involving matter and gauge fields is more complex. 
From Eq.~\eqref{eq:Hamiltonian} it is evident that $g^{-2}$ only acts as a scaling factor for the individual terms contributing to the expectation value $\langle\hat{H}(\boldsymbol{\theta})\rangle$. Hence, the cost function can be evaluated simultaneously for all values of $g^{-2}$ after each measurement. This allows us to generate a data storage system and pair it with the BO algorithm. We allow the algorithm to initialize the storage by taking 130 measurements, before sampling up to 300 times or until convergence is achieved. During the optimization of the parameters for a single value of $g^{-2}_{i}$, the outcomes are processed and saved for successive searches. After successful optimization for $g^{-2}_{i}$, this process is repeated for $g^{-2}_{i+1}$, considering all previous outcomes. With the assumption, that the optimal solutions change smoothly for small variations in the bare coupling, the already found minimum for $g^{-2}_i$ is a good candidate for an initial guess for $g^{-2}_{i+1}$. Consequently, the algorithm now does not require sampling the whole parameter space but can invest in refining the search around promising solutions. As the bare coupling is changed, the available \textit{knowledge} increases, in turn yielding faster convergence.

We also modify a recently proposed trust region BO \cite{Eriksson2019Scalable} to accept noisy cost function evaluations, couple it to the data storage, and perform two sweeps in the coupling ranging from $g^{-2} = 0.01$ to $g^{-2} = 100$. The best values are obtained after the last run. 

%
\subsection{VQE Measurements for Qudits}
\label{sec:measurement}
In the following, we explain the qudit measurements and basis decompositions that are performed in our VQE experiments.
Contrary to qubits where one naturally chooses the Pauli product basis, there is no clear indication for the best possible basis to decompose a given qudit (possibly mixed-dimensional) Hamiltonian. 
This is due to the fact that the Pauli operators allow for an efficient classical determination of a circuit that diagonalizes a group of commuting Pauli operators so they can be measured simultaneously~\cite{Shlosberg2023adaptiveestimation}.

For qudits, we represent the Hamiltonian in terms of the so-called clock and shift operators, which are normalized by the generalized Clifford group comprising of the $d$-dimensional SUM (generalized CNOT), Hadamard and S gates~\cite{gottesman1999fault,jena_future}. More precisely, the clock and shift operators are respectively defined as $\hat{Z} = \mathrm{diag}\lbrace e^{\frac{2 \pi i}{d}k}\rbrace_{k=0}^{d-1}$ and $(\hat{X})_{kl}=\delta_{k,j-1} + \delta_{k,d}\delta_{j,d}$.
By expressing the matter and gauge fields, as well as the operators $\hat{P}_{\vec{n}}$ as linear combinations of tensor products of $\hat{X}^\alpha \hat{Z}^\beta$ ($\alpha,\beta\in \{0,\dots,d-1 \}$), we can therefore write the Hamiltonian $\hat{H}$ with contributions from Eqs.~\eqref{eq:elec_rep} and~\eqref{eq:magn_rep} in terms of the clock and shift operators only, $\hat{H} = \sum_i c_i \hat{O}_i$,~\cite{jena_future}.
Here, $c_i \in \mathbb{C}$ and $\hat{O}_i = \bigotimes_{j=1}^m \hat{X}^{\alpha_j^i} \hat{Z}^{\beta_j^i}$ for $m$ qudits and coefficients $\alpha_j^i$ and $\beta_j^i$ determined by the decomposition of $\hat{H}$. 

Normalizing these operators with gates from the generalized Clifford group then maps a subset of the $\hat{O}_i$ operators to clock operators $\hat{Z}^{\gamma_i}$ (with all $\gamma_i$ not necessarily identical), which can subsequently be measured simultaneously.

\section{Real-time Evolution}
\label{sec:realtime}
In this section we show how to use the qudit tools that we developed in the context of equilibrium problems (see Fig.~\ref{fig:1} and Fig.~\ref{fig:2}) to study time evolutions in LGTs, an area that is inaccessible to MCMC methods due to sign-problems~\cite{Troyer:2004ge, Gattringer:2016kco}. As proof-of-concept demonstration, we consider here a single plaquette with open boundary conditions, as in Sec.~\ref{sec:open} of the main text. As in the case of the VQE demonstration, we study this model with a hybrid qubit-qutrit system, which can be realized within the same trapped-ion chain. As initial state of the time evolution we choose the bare vacuum of the system $|vvvv, 0⟩$, where no particles (first four entries) or gauge field excitations (last entry) are present. We study its time evolution under the Hamiltonian given in Eq.~\eqref{eq:openplaq_qubit}. As depicted in \figref{fig:encoding_si}, the initial state in qubit-qudit form is given by $\ket{\downarrow\uparrow\downarrow\uparrow,0}$. This time evolution can be interpreted as a quench from the strong coupling regime ($g^{-2} \ll 1$), where the bare vacuum is the ground state, to a weaker coupling. To simulate this time evolution on our quantum hardware, we perform a Trotter protocol~\cite{trotter_product_1959,Lloyd1996}. We decompose the Hamiltonian into a sum $\hat{H} = \sum_{i} \hat{h}_{i}$, then the state at time $t$ calculated with $N_T$ Trotter steps is given by
	\begin{equation}
		\underbrace{\prod_{i=1} \exp\left( -i \frac{t}{N_T} \hat{h}_{i} \right) \dots \prod_{i=1} \exp\left( -i \frac{t}{N_T} \hat{h}_{i} \right)  }_{N_T \text{times}} \ket{\downarrow\uparrow\downarrow\uparrow,0}.
	\end{equation}
	We choose the following decomposition into Trotter steps for the Hamiltonian given in Eq.~\eqref{eq:openplaq_qubit}:
	\begin{subequations}
		\label{eq:trotter_decomposition}
		\begin{align}
			\hat{h}_{1} &= \Omega\left( \hat{\sigma}^{+}_{1} \hat{U}^\dagger \hat{\sigma}^{-}_{2} -  \hat{\sigma}^{+}_{4} \hat{\sigma}^{-}_{3} + \text{H.c.} \right), \\
			\hat{h}_{2} &= \Omega \left( \hat{\sigma}^{+}_{2} \hat{\sigma}^{-}_{3} - \hat{\sigma}^{+}_{1} \hat{\sigma}^{-}_{4} +\text{H.c.} \right),\\
			\hat{h}_{3} &= \frac{g^{2}}{4} \left( \hat{\sigma}^{z}_{1} \hat{\sigma}^{z}_{4} + - 4\hat{E} \hat{\sigma}^{z}_{1}+  2\hat{E} \hat{\sigma}^{z}_{2} - 2\hat{E} \hat{\sigma}^{z}_{4}\right),\\
			\begin{split}
				\hat{h}_{4} &= \frac{m}{2} \left( \hat{\sigma}^{z}_{1} - \hat{\sigma}^{z}_{2} + \hat{\sigma}^{z}_{3} - \hat{\sigma}^{z}_{4} \right) + \\
				&\quad \frac{g^{2}}{4} \left(  8 \hat{E}^{2} - 4\hat{E}  + \hat{\sigma}^{z}_{1} - \hat{\sigma}^{z}_{2} + 3\right),
			\end{split}\\
			\hat{h}_{5} &= -\frac{1}{g^{2}} \left( \hat{U} + \hat{U}^{\dagger} \right).
		\end{align}
	\end{subequations}

\begin{figure}
	\centering
	\includegraphics[width=0.9\columnwidth]{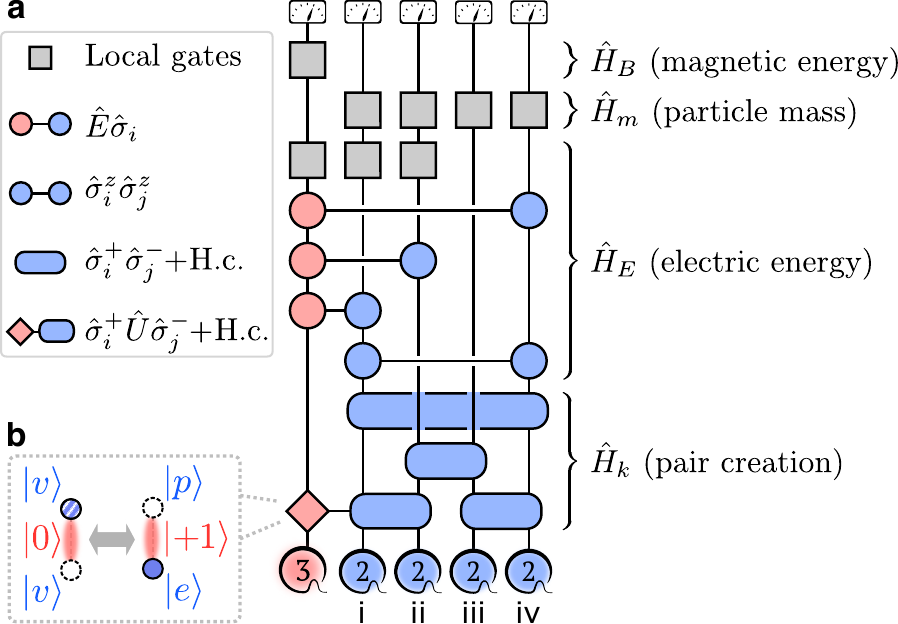}
	\caption{\textbf{Time evolution circuit for 2D-QED with matter.} \textbf{a}, Circuit for the Trotter time evolution of a plaquette with dynamical matter, following the decomposition given in Eqs.~\ref{eq:trotter_decomposition}. For details on the model see \methods~\ref{sec:realtime}. \textbf{b}, An example of the transitions induced by the kinetic term of the Hamiltonian.}
	\label{fig:exp_trotter}
\end{figure}

We experimentally realize the above Trotter protocol for a single Trotter step $N_T=1$. The gates appearing in the Trotter circuit (see Fig.~\ref{fig:exp_trotter}a) are experimentally implemented as follows: All local gates, including $\hat{\sigma}^{z}$, $\hat{U}$, and $\hat{E}^{(2)}$ are realized as in the main text. Terms of the form $\hat{E} \hat{\sigma}^{z}$ are (for qutrits) effectively equivalent to $\hat{\sigma}^{z}\hat{\sigma}^{z}$, and are realized by a standard M\o lmer-S\o rensen (MS) gate~\cite{sorensen1999quantum} that is locally rotated using (subspace) Hadamard gates on all involved ions. Gates of the form $\hat{\sigma}^{+} \hat{\sigma}^{-} + \text{H.c.}$ are realized as $\hat{\sigma}^{x}\hat{\sigma}^{x} + \hat{\sigma}^{y}\hat{\sigma}^{y}$, which corresponds to a sequence of two locally rotated MS gates. This leaves only the term $\hat{\sigma}^{+}_{1} \hat{U}^\dagger \hat{\sigma}^{-}_{2} + \text{H.c.}$. Extending the decomposition given in Ref.~\cite{andrade_engineering_2022}, this three-body coupling can be realized using 24 two-body MS gates:

	\begin{align}
		e^{i \theta (\hat{\sigma}^{+}_{i} \hat{U}^\dagger_{j} \hat{\sigma}^{-}_{k} + \text{H.c.})} =& 
		e^{-\frac{i\pi}{4}\hat{\sigma}^{y}_{i}\hat{\sigma}^{z_{01}}_{j}} 
		e^{\frac{i\theta}{4}\hat{\sigma}^{x_{01}}_{j}\hat{\sigma}^{x}_{k}}
		e^{\frac{i\pi}{4}\hat{\sigma}^{y}_{i}\hat{\sigma}^{z_{01}}_{j}} \nonumber \\
		&e^{-\frac{i\pi}{4}\hat{\sigma}^{y}_{i}\hat{\sigma}^{y_{01}}_{j}} 
		e^{\frac{i\theta}{4}\hat{\sigma}^{z_{01}}_{j}\hat{\sigma}^{y}_{k}}
		e^{\frac{i\pi}{4}\hat{\sigma}^{y}_{i}\hat{\sigma}^{y_{01}}_{j}} \nonumber \\
		&e^{-\frac{i\pi}{4}\hat{\sigma}^{x}_{i}\hat{\sigma}^{y_{01}}_{j}} 
		e^{-\frac{i\theta}{4}\hat{\sigma}^{z_{01}}_{j}\hat{\sigma}^{x}_{k}}
		e^{\frac{i\pi}{4}\hat{\sigma}^{x}_{i}\hat{\sigma}^{y_{01}}_{j}} \nonumber \\
		&e^{-\frac{i\pi}{4}\hat{\sigma}^{x}_{i}\hat{\sigma}^{x_{01}}_{j}} 
		e^{-\frac{i\theta}{4}\hat{\sigma}^{z_{01}}_{j}\hat{\sigma}^{y}_{k}}
		e^{\frac{i\pi}{4}\hat{\sigma}^{x}_{i}\hat{\sigma}^{x_{01}}_{j}} \nonumber \\
		&e^{-\frac{i\pi}{4}\hat{\sigma}^{y}_{i}\hat{\sigma}^{z_{12}}_{j}} 
		e^{\frac{i\theta}{4}\hat{\sigma}^{x_{12}}_{j}\hat{\sigma}^{x}_{k}}
		e^{\frac{i\pi}{4}\hat{\sigma}^{y}_{i}\hat{\sigma}^{z_{12}}_{j}} \nonumber \\
		&e^{-\frac{i\pi}{4}\hat{\sigma}^{y}_{i}\hat{\sigma}^{y_{12}}_{j}} 
		e^{\frac{i\theta}{4}\hat{\sigma}^{z_{12}}_{j}\hat{\sigma}^{y}_{k}}
		e^{\frac{i\pi}{4}\hat{\sigma}^{y}_{i}\hat{\sigma}^{y_{12}}_{j}} \nonumber \\
		&e^{-\frac{i\pi}{4}\hat{\sigma}^{x}_{i}\hat{\sigma}^{y_{12}}_{j}} 
		e^{-\frac{i\theta}{4}\hat{\sigma}^{z_{12}}_{j}\hat{\sigma}^{x}_{k}}
		e^{\frac{i\pi}{4}\hat{\sigma}^{x}_{i}\hat{\sigma}^{y_{12}}_{j}} \nonumber \\
		&e^{-\frac{i\pi}{4}\hat{\sigma}^{x}_{i}\hat{\sigma}^{x_{12}}_{j}} 
		e^{-\frac{i\theta}{4}\hat{\sigma}^{z_{12}}_{j}\hat{\sigma}^{y}_{k}}
		e^{\frac{i\pi}{4}\hat{\sigma}^{x}_{i}\hat{\sigma}^{x_{12}}_{j}},
\end{align}
where $\hat{\sigma}^{z_{ij}}$ refers to the Pauli-Z operator acting on the subspace spanned by the states $\{\ket{i},\ket{j}\}$. In our simple proof-of-principle experiment, for a given initial state, it is not necessary to implement the full coupling term. Instead, we only implemente the non-trivial component for the given initial state, requiring just 6 MS gates. Notably, all required interactions for the mixed-dimensional time evolution of 2D QED are already part of our toolbox in the main text.

	The minimal realization using one Trotter step allows us already to observe the dynamics of the 
	mean particle number density $\nu = \frac{\expval{\hat{H}_{m}}}{4} + \frac{1}{2}$, as shown in Fig.~\ref{fig:exp_trotter}. Note that $\nu=0$ for the bare vacuum, and $\nu=1$ for a completely filled system. We can see that immediately after the quench, the kinetic part of the Hamiltonian brings pair-creation processes into play, therefore increasing the particle density of the state. After some particles and antiparticles are created, however, pairs can be annihilated, leading to a decrease in the average particle density. Fig.~\ref{fig:long_trotter} shows how the dynamics of the particle number density $\nu$ and the plaquette expectation value $\plaqop$ is approximated by the Trotter time evolution, with $N_T = 1$, $N_T = 10$, and $N_T \rightarrow \infty $ (exact result).

\section{Quantum Computing for Particle Physics}
\label{Sec:Roadmap}

Predictions in particle physics that involve the strong force can in general not be performed with perturbative methods and require therefore computer simulations. So far, lattice gauge theory (LGT) is the only general solution available in this endeavor. The standard approach of performing LGT simulations using the action formalism and Markov Chain Monte Carlo (MCMC) calculations has proven to be extremely successful but breaks down in situations that involve, for example, finite fermionic chemical potentials (i.e.\ finite matter densities) and real-time dynamics~\cite{Gattringer:2016kco}. This limitation motivates an intensive ongoing search for new computational methods. Since quantum computing does not face the sign-problems that currently impede MCMC-based approaches, LGT quantum simulations offer an exciting opportunity to explore new avenues for gauge theory calculations in high energy physics~\cite{dimeglio2023quantum,beck2023quantum,bauer_quantum_2023}. In particular, quantum computers could allow making inroads into the two big problem areas outlined below.

\medskip
\textbf{Equilibrium physics} -- A concrete example of a problem that is out of reach for current MCMC-based methods is the phase diagram of quantum chromodynmics (QCD), which describes quarks and gluons at different fermionic chemical potentials and temperatures~\cite{bogdanov_snowmass_2022}. Forming the basis for understanding which states of matter are possible in Nature, the QCD phase diagram informs important open questions in nuclear physics, particle physics and cosmology, and could be explored by future LGT quantum simulations at finite densities.

\medskip
\textbf{Dynamics} -- Another specific example of a problem area that is currently inaccessible to MCMC calculations is the real-time evolution in collider physics~\cite{banuls2019simulating}. Since there is no fundamental roadblock to study dynamics with quantum computers, LGT quantum simulations could open the door to a better understanding of particle collisions, i.e. the modelling and simulation of how the different degrees of freedom evolve in time during scattering processes observed at collider experiments at CERN and other facilities.

\medskip
Traditional methods cannot even begin addressing the problem areas outlined above, as MCMC calculations do not converge and the method is unsuitable from the outset. Using quantum computers as a new tool for investigating the above examples, will involve a number of challenges including the following two key steps: 
\begin{enumerate}
		\item Simulating LGTs in 3D, which requires representing gauge degrees of freedom with three and more levels.
		\item Simulating LGTs with non-Abelian symmetry groups, i.e.\ including color-degrees of freedom.
\end{enumerate}

\section{Qudit Advantage} \label{Sec:Supp:Advantage}
We are now comparing our native qudit implementation to a standard \textit{qubit} implementation of the circuits presented in the main text. Here we follow the efficient paradigm outlined in Ref.~\cite{paper1}, where high-dimensional gauge fields are encoded into qubits using a one-hot encoding: we map each $d$-level gauge field onto $d$ qubits. All qubits are in the up state, except for one; which qubit is in the down state indicates which of the $d$ levels is occupied. This has the advantage of greatly simplifying the required entangling gates and thus reducing the circuit depth, at the cost of an increase in qubit number. One may also consider a binary encoding, where we map a qudit onto $\lceil{\log_2 d}\rceil$ qubits. This optimizes the required register size at the cost of more complicated entangling gates. For the qudit platform, we make the assumption that an embedded CNOT gate (i.e.\ a CNOT gate acting on a qubit subspace of the qudit Hilbert space) can be performed with roughly equal fidelity as the same gate in a pure qubit system. This is the case for the trapped-ion platform used here. 

\begin{figure}[ht]
	\centering
	\includegraphics[width=0.5\columnwidth]{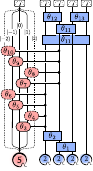}
	\caption{Variational ansatz for simulating a single plaquette with open boundary conditions when the gauge field is represented by a ququint. The gates used in the circuit are defined in \methods~\ref{sec:circuit_gates}.}
	\label{fig:open_ququint}
\end{figure}

We now consider the scaling of circuit complexity with increasing gauge-field truncation to capture the physics more accurately. Starting with the full gauge theory of one plaquette in 2D, as shown in the main text Fig.~\ref{fig:1}, we propose a physically motivated generalization of the VQE circuit, which requires a linear increase in gate count with gauge field dimension. In particular, we show in \figref{fig:open_ququint} the generalization of the circuit when using a ququint truncation for the gauge field. More generally the qudit implementation of the circuit requires three CNOT for each of the qubit $A$ gates, and two CNOT for each of the C-ROT for a total of $18+4(d-1)$ CNOT gates. In the case of the qubit implementation studied in Ref.~\cite{paulson_simulating_2021} $18+36(d-1)$, since the implementation of a control-rotation, in this case, requires six CNOT. This shows that the scaling of the qudit platform is about an order of magnitude better than a comparable qubit platform, already in this simplest instance of the problem. Note that we use the CNOT-based gate count for cross-platform comparability, while our experiment is using a different decomposition of the gates, specific to trapped ions. The results are summarized in Tab.~\ref{tab:open}.

\begin{table}[htb]
	\centering
	\begin{tabularx}{\columnwidth}{c || *3{>{\centering\arraybackslash}X|} | *3{>{\centering\arraybackslash}X|} } \toprule
		& \multicolumn{3}{c||}{Qudit encoding} & \multicolumn{3}{c|} {Qubit encoding} \\ \hline \hline
		Dimension $d$ & 3 & 5 & 7 & 3 & 5 & 7 \\
		Register size & 5 & 5 & 5 & 7 & 9 & 11 \\
		CNOT count & 26 & 34 & 42 & 90 & 162 & 234 \\ 
		CNOT fidelity &  \multicolumn{6}{c|}{99\%}\\
		Approx. circ. fid. & 77\% & 71\% & 66\% & 40\% & 20\% & 10\% \\ \hline
		CNOT fidelity &  \multicolumn{6}{c|}{99.5\%}\\
		Approx. circ. fid. & 88\% & 84\% & 81\% & 64\% & 44\% & 31\% \\ \hline
	\end{tabularx}
	\caption{\textbf{Circuit complexity of a single plaquette of the full gauge theory in 2D}. The complexity is measured in a hardware-agnostic way by the gate count when all entangling gates (qubit and qudit) are decomposed into two-level CNOT gates under the assumption that these gates work equally well in qudit as in qubit systems. More efficient decompositions might exist by using custom hardware-efficient gates.}
	\label{tab:open}
\end{table}

The results are even more striking in the case of a pure gauge theory as discussed in Fig.~\ref{fig:2} of the main text. Here, the qudit platform shows no scaling with gauge field truncation, as the added complexity is only in the local qudit operations. In the qubit platform, on the other hand, the gate count scales as $48+12d$, and the register size scales as $3d$, see Tab.~\ref{tab:periodic}. Considering an encoding optimized for register size, this growth can be reduced to $3\lceil\log_2(d)\rceil$, at the cost of a gate count that scales in general quadratically with the dimension.

\begin{table}[ht]
	\centering
	\begin{tabularx}{\columnwidth}{c || *3{>{\centering\arraybackslash}X|} | *3{>{\centering\arraybackslash}X|} } \toprule
		& \multicolumn{3}{c||}{Qudit encoding} & \multicolumn{3}{c|} {Qubit encoding} \\ \hline
		Dimension $d$ & 3 & 5 & 7 & 3 & 5 & 7 \\
		Register size & 3 & 3 & 3 & 9 & 15 & 21 \\
		CNOT count & 8 & 8 & 8 & 84 & 108 & 132 \\ 
		CNOT fidelity &  \multicolumn{6}{c|}{99\%}\\
		Approx. circ. fid. & 92\% & 92\% & 92\% & 43\% & 34\% & 27\% \\ \hline
		CNOT fidelity &  \multicolumn{6}{c|}{99.5\%}\\
		Approx. circ. fid. & 96\% & 96\% & 96\% & 66\% & 58\% & 52\% \\ \hline
	\end{tabularx}
	\caption{\textbf{Circuit complexity of a pure gauge plaquette with periodic boundary conditions}. The complexity is measured in a hardware-agnostic way by the gate count when all entangling gates (qubit and qudit) are decomposed into two-level CNOT gates under the assumption that these gates work equally well in qudit as in qubit systems. More efficient decompositions might exist by using custom hardware-efficient gates.}
	\label{tab:periodic}
\end{table}

Finally, one might suspect that the price to pay for the reduced circuit complexity in the qudit approach is a less favorable runtime. While it is generally true that the qudit circuits take longer due to long readout times, we show in Tab.~\ref{tab:runtimes} that even the tradeoff between qudit readout ($(d-2)4$ ms) and qubit gate complexity ($0.2$ ms per gate) is slightly in favor of the qudit system for the problems studied here.
\begin{table}[ht]
	\centering
	\begin{tabularx}{\columnwidth}{c || *3{>{\centering\arraybackslash}X|} | *3{>{\centering\arraybackslash}X|} } \toprule
		& \multicolumn{3}{c||}{Qudit encoding} & \multicolumn{3}{c|} {Qubit encoding} \\ \hline
		Dimension $d$ & 3 & 5 & 7 & 3 & 5 & 7 \\
		Register size & 3 & 3 & 3 & 9 & 15 & 21 \\
		Rel. runtime open & 1 & 1.52 & 2.04 & 1.48 & 2.26 & 3.04 \\ 
		Rel. runtime periodic & 1 & 1.54 & 2.08 & 1.76 & 2.08 & 2.41 
	\end{tabularx}
	\caption{\textbf{Runtimes of one execution of the qubit and qudit implementations}. These times are calculated for the device used in the main text and may differ in other architectures.}  
	\label{tab:runtimes}
\end{table}

\section{VQE Circuit for Pure-Gauge QED with Periodic Boundary Conditions}
\label{sec:periodic_circuit}
We explain here the essential points of the design of the circuit used to study the pure-gauge QED model producing the results shown in \figrefp{fig:2}{d}. A more detailed discussion is found in Ref.~\cite{paulson_simulating_2021}. 

As for the case of 2D-QED with matter discussed in the main text, the form of the circuit is inspired by the Hamiltonian that the VQE solves, providing us with a scalable strategy that works also for higher truncations of the gauge field, and also with two different variational circuits for the electric and magnetic representation respectively. Let us consider the electric Hamiltonian given in Eqs.~\eqref{eq:elec_rep}, its ground state is found via the variational ansatz given by the circuit in~\figrefp{fig:2}{c}. After creating superpositions of all states for rotator $\hat{R}_2$, there is a layer of C-ROT gates that reflects the coupling between rotators given by $\hat{H}^{(e)}_{E}$, where we have used the symmetry between rotator $\hat{R}_1$ and $\hat{R}_3$ to reduce the number of independent variational parameters. For larger values of $g^{-2}$, the system tends to occupy states where the three rotators occupy the same electric field eigenstate, therefore we spread the population of rotators 1 and 3 controlling suitably on the corresponding state of rotator 2.

\begin{figure}[ht]
	\centering
	\includegraphics[scale=1.1]{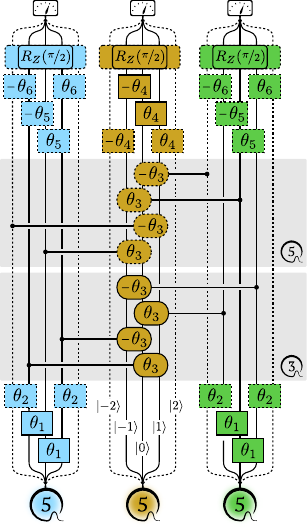}
	\caption{Variational circuit for the magnetic representation for the qutrit (solid lines) and the ququint (all, except box marked with qutrit) truncation. The explicit form of the gates employed is given in \methods~\ref{sec:circuit_gates}. The last gate in the circuit is defined as $\hat{R}_Z(\theta) = \exp\left(-i\theta \hat{Z} \right)$, where $\hat{Z} = \operatorname{diag}(-1,0,1)$ for the qutrit, and $\hat{Z} = \operatorname{diag}(-2,-1,0,1,2)$ for the ququint truncation.}
	\label{fig:periodic_magnetic_circuit}
\end{figure}

The circuit for the magnetic Hamiltonian given in Eqs.~\eqref{eq:magn_rep} and depicted in Fig.~\ref{fig:periodic_magnetic_circuit} is designed with similar arguments.

\section{Experimental Details}
\label{sec:more_experimental_details}

\subsection{Optimal Encoding and Cross-Talk Suppression}
\begin{figure}[ht]
	\centering
	\includegraphics[scale=1.]{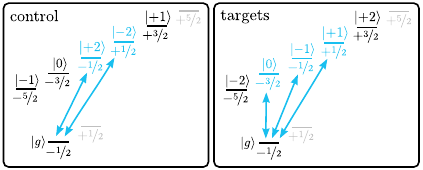}
	\caption{Optimal encodings of control and target qudits in the $D_{\nicefrac{5}{2}}$ manifold of $^{40}\mathrm{Ca}^+$ applied for the pure gauge ququint circuit in~\figrefp{fig:2}{c} -- the encoding is chosen such that relevant states for the C-ROTs have the strongest coupling to the blue sideband. We seek to minimize phase errors by allowing only $\pi$ rotations on all transitions, which induce large Stark shifts on the spectator levels.}
	\label{fig:exp_optimal_encoding}
\end{figure}

Employing embedded C-ROTs requires careful treatment of the frequency spectrum and the structure of the quantum circuit to optimize the encoding of qudits in trapped ions. From the pure gauge variational ququint circuit shown in~\figrefp{fig:2}{c}, it is evident that only few states participate in the C-ROTs, namely the states $\ket{\pm 2}$ of control ququint and $\ket{0}$, $\ket{\pm 1}$ of the two target qudits. Optimal encoding yields the strongest coupling of the qudit states with the auxiliary ground state $\ket{g}$ on the blue sideband while simultaneously minimizing accumulating phase errors as a consequence of imperfect AC Stark shift compensation. The latter is achieved by restricting transitions, which lead to large Stark shifts, to static angles of $\pm \pi$, making the shift constant and easier to compensate. Figure~\ref{fig:exp_optimal_encoding} shows the chosen encodings for control and target qudits for the ququint circuit shown in~\figrefp{fig:2}{c} for the electric representation. The states with the largest coupling to $\ket{g}$ are highlighted in blue. Note, that in the magnetic representation, a different encoding is chosen. 

\begin{figure}[ht]
	\centering
	\includegraphics[scale=1.]{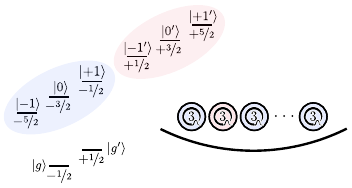}
	\caption{Next-neighbor cross-talk suppression in qutrits by proper choice of encoding -- spatially neighboring ions encode the qutrit in different sets of Zeeman states with each set separated by a few $\mathrm{MHz}$. This encoding pattern is well suited to suppress cross-talk~\cite{Meth2022}, allowing higher axial trap frequencies, reducing the inter-ion distance.}
	\label{fig:exp_qutrit_even_odd}
\end{figure}

\begin{figure}[h]
	\centering
	\includegraphics[scale=1.1]{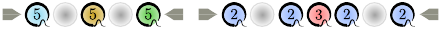}
	\caption{Artificially increasing the inter-qudit distance by introducing \textit{buffer} ions. Buffer ions do not contribute to the circuit and do not couple with the light fields used to control the qudits. This strategy is employed for the $d=5$ pure gauge system, as well as for the circuits describing dynamical matter. For the latter case, we also chose to re-order the qudits inside the ion string to enhance the performance of the M\o lmer-S\o rensen entangling gates on the qubits.}
	\label{fig:exp_buffer_ions}
\end{figure}

The pure-gauge qutrit circuit can be constructed in such a way that spatially neighboring ions store the qutrit in different sets of Zeeman levels, with each set separated by frequencies of a few $\mathrm{MHz}$. Assigning integers to each ions in the string from left to right, ions with even indices encode the qutrit in Zeeman states with positive magnetic quantum numbers $\ket{-1'} = D_{+\nicefrac{1}{2}}$, $\ket{0'} = D_{+\nicefrac{3}{2}}$ and $\ket{+1'} = D_{+\nicefrac{5}{2}}$, coupled via the second ground state $\ket{g'} = S_{+\nicefrac{1}{2}}$ as shown in Fig.~\ref{fig:exp_qutrit_even_odd}. With this spectroscopic decoupling scheme, next-neighbor cross-talk is effectively eliminated~\cite{Meth2022} and higher axial trap frequencies can be used. However, spectroscopic decoupling is not possible for ququints, as no bi-partition of the $D_{\nicefrac{5}{2}}$ manifold exists in $^{40}\mathrm{Ca}^+$ for qudits of dimension $d > 3$; here one would require different ion species or isotopes to apply this scheme. 
A second approach to reduce next-neighbor cross-talk is to increase the spatial separation between ions encoding the qudits. This can be achieved in two ways, either lowering the axial confinement by reducing the end-cap voltage of the ion trap or by introducing additional \textit{buffer} ions, which are transferred to a \textit{hidden} state which does not couple with the laser pulses driving the actual quantum circuit. The former strategy is less suited for our specific circuits due to significant \textit{crowding} of the mode spectrum, making it difficult to faithfully implement the C-ROT operations. Thus, for the $d=5$ pure gauge circuit as well as the gauge-matter circuit we chose the second method and introduced a pair of extra ions - in the pure gauge circuits, these buffers are located between each qudit, while the open plaquette circuit requires a buffer between each of the matter qubits, while we order the ion string in such a way that the gauge field qudit is physically located in the center of the string as shown in \figref{fig:exp_buffer_ions}. This re-ordering ensures optimal performance of the M\o lmer-S\o rensen gates on the matter qubits at the very end of the circuit. 
Additionally to the two approaches presented in this section, we also employ composite rotations \cite{Meth2022} when transferring the buffer ions to their hidden state. 
\subsection{Gate Count Optimization}
The structure of the quantum circuits enables us to implement shortcuts, reducing the number of entangling gates; these shortcuts are best explained by the circuit in~\figrefp{fig:1}{d}. First, if the input state of one of the qudits is known, the number of interactions with the phonon mode $B(\theta, \phi)$ can be reduced to $n_\mathrm{B(\theta, \phi)} = 2$ compared to the general case of a C-ROT being composed of $n_{B(\theta, \phi)} = 5$ BSB pulses. This strategy is applied to the matter \textit{qubits} in the gauge-matter-circuit for the angles $\theta_1$, $\theta_2$ and $\theta_7$, where the state of the \textit{`left'} qubit entering each operation is exactly known. Second, if two C-ROTs are executed successively on the same control state as it would be for the pairs $(\theta_3, \theta_4)$ and $(\theta_5, \theta_6)$ in~\figrefp{fig:1}{d}, two successive $B(\theta,\phi)$ cancel out as it can be imagined by extending the sequence in Fig.~\ref{fig:exp_crot_sequence} by a second C-ROT. Third, we implement the SWAP operations with angles $\theta_7$, $\theta_8$ and $\theta_9$ by sequences of three M\o lmer-S\o rensen (MS) gates~\cite{sorensen1999quantum} and local single-qubit rotations, which, in this case, is more favorable than the blue-sideband approach in terms of gate count.

\subsection{Measurement and Readout of Qudits}
\begin{figure}[ht]
	\centering
	\includegraphics[scale=1]{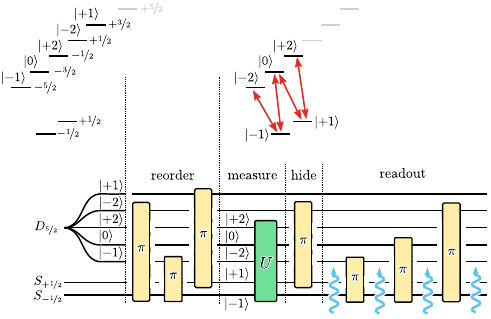}
	\caption{Measurement and readout of ququints -- the ququint is initially encoded in the $D_{\nicefrac{5}{2}}$ manifold of $^{40}\mathrm{Ca}^+$ similar to Fig.~\ref{fig:exp_optimal_encoding} (left). It is then reordered by a sequence of $\pi$-pulses, such that the states are arranged in the shape of the letter W (right). The coupling between states $\ket{i\pm 1}\leftrightarrow\ket{i}$ is required to faithfully implement the measurement operation $U$ in green. After hiding $\ket{+1}$ in the $D_{\nicefrac{5}{2}}$ manifold, the readout sequence is executed as a series of detection and re-cooling \& optical pumping pulses indicated by blue arrows. Additional $\pi$-pulses rotate the desired state to the ground state manifold, such that photons can be scattered on the short-lived $S_{\nicefrac{1}{2}}\leftrightarrow P_{\nicefrac{1}{2}}$ transition (see \figref{fig:exp_ca40_scheme}).}
	\label{fig:exp_measurement_scheme}
\end{figure}

Measurements on a qudit with dimension $d$ are implemented by a series of single-qudit operations followed by a sequence of $d-1$ detection events and local $\pi$ rotations. Our optimal encoding of the qudits requires an additional reordering of the quantum states to match a $W$-pattern as shown in Fig.~\ref{fig:exp_measurement_scheme}. In this arrangement any state $\ket{i}$ directly couples with $\ket{i\pm 1}$, which enables a straightforward QR-decomposition-based compilation of the measurement operator into a sequence of single qudit rotations~\cite{Ringbauer2022A-universal}.
After the measurement operation has been applied, the readout is implemented by a sequence of $d-2$ local $\pi$-flips and $d-1$ detection events. Each event is composed of a $1~\mathrm{ms}$ window, during which the fluorescence of each ion is recorded on a sensitive EMCCD camera, followed by $2~\mathrm{ms}$ of Doppler cooling and $0.5~\mathrm{ms}$ of polarization gradient cooling; this re-cooling pattern is not required after the last detection. We mention, that the duration of the readout in our experiment scales as $\approx (d-2)\cdot 3.5~\mathrm{ms} + 1~\mathrm{ms}$ for any qudit dimension $d$ -- this is on the order of the duration of the ground state cooling sequence at the beginning of each experiment run.

\section{Noise Model}
\label{sec:noise_model}
While systems consisting of qubits only can often be accurately described by generic noise models such as simple depolarising noise, we found that for our qudit systems this is too reductive. Therefore, we construct a more realistic noise model based on the physics of how the gates of our circuits are realized. In particular, our heuristic model includes imperfect precision in the variational parameters and dephasing of the prepared quantum state. We find good agreement between the experimental results and the simulated data.

\textbf{Amplitude Fluctuations. } 
We expect that the precision in the gate angles we aim to apply throughout the experiment suffers from fluctuations of the addressing laser amplitude. A gate $\hat{U}(\theta)$ is hence replaced by $\hat{U}(\theta + \delta \theta)$, where $\delta\theta$ is a stochastic variable extracted from a Gaussian $\mathcal{N}(0,\sigma_{\text{a}}^2)$ with mean $\mu=0$ and variance $\sigma_{\text{a}}^2$. Given the higher fidelity with which single qudit gates can be realized compared to entangling gates, we use in our model a variance $\sigma_{\text{a}}^2$ for parametrized entangling gate, and a variance $\sigma_{\text{a}}^2/10$ for single qudit gates.

\textbf{Phase Errors. } 
In trapped ion systems, while single qudit operations can be carried out with a fidelity close to unity, two-qudit entangling gates and controlled excitation of the ions' motion are the dominant sources of errors. C-ROT operations in qudits introduce phase noise due to AC Stark shifts of different magnitude (see Methods~\ref{sec:experimetal_details}): (1) exciting the motion on a chosen control state $\ket{i}$ introduces a (small) phase error $\phi_i$ on that state due to imperfect compensation and (2) spectator states that are not involved in the interaction ($\ket{j}$, with $j\neq i$) accumulate a phase $\Phi_j$ due to imperfect compensation of AC stark shifts. Each C-ROT is a composition of single qudit operations as shown in Fig.~\ref{fig:exp_crot_sequence}---for any blue sideband operation exciting the motion via the state $\ket{i}$, we identify the mapping $c_i \ket{i} + \sum_{j\neq i} c_j \ket{j} \mapsto e^{i\phi_i}c_i\ket{i} + \sum_{j\neq i}e^{i\Phi_j}c_j\ket{j}$. As can be deduced by counting how many times each level in Fig.~\ref{fig:exp_crot_sequence} is involved in a sideband pulse, the control qudit is mapped to the state in Eq.~\eqref{eq:app_phase_errors_control}, while the target acquires phases according to Eq.~\eqref{eq:app_phase_errors_target}.

\begin{equation}
	c_i \ket{i} + \sum_{j\neq i} c_j\ket{j} \mapsto e^{i2\phi_i} c_i\ket{i} + \sum_{j\neq i} e^{i2\Phi_j}c_j\ket{j},
	\label{eq:app_phase_errors_control}
\end{equation}

\begin{equation}
	\begin{split}
		&c_i\ket{i} + c_j\ket{j} + \sum_{k\neq i, j} c_k\ket{k} \mapsto \\
		&e^{i(2\phi_i + \Phi_i)}c_i\ket{i} + e^{i(2\Phi_j + \phi_j)}c_j\ket{j} + \sum_{k\neq i, j} c_k e^{i3\Phi_k}\ket{k}.
		\label{eq:app_phase_errors_target}
	\end{split}
\end{equation}
The second-beam technique~\cite{haffner2008quantum} applied to cancel the AC Stark shift on the \emph{actively} driven transition $\ket{g}\leftrightarrow\ket{i}$ is robust to intensity noise, as both beams are affected equally by fluctuations. Thus, the phase error $\phi_i$ is considered to be small compared to $\Phi_j$ and can be neglected for simplicity. Furthermore, we make the simplifying assumption that the passively (in software) compensated spectator shifts yield the same phase error for all levels $\Phi_j\mapsto \Phi$, and thus the mappings simplify to Eq.~\eqref{eq:app_phase_errors_control_phi} and Eq.~\eqref{eq:app_phase_errors_target_phi}.
\begin{equation}
	c_i\ket{i} + \sum_{j\neq i} c_j\ket{j} \mapsto c_i\ket{i} + e^{i2\Phi}\sum_{j\neq i}c_j\ket{j},
	\label{eq:app_phase_errors_control_phi}
\end{equation}

\begin{equation}
	\begin{split}
		&c_i\ket{i} + c_j\ket{j} + \sum_{k\neq i, j} c_k\ket{k} \mapsto \\&e^{i\Phi}c_i\ket{i} + e^{i2\Phi}c_j\ket{j} + e^{i3\Phi}\sum_{k\neq i, j} c_k\ket{k}.
		\label{eq:app_phase_errors_target_phi}
	\end{split}
\end{equation}
Consecutive C-ROTs controlled by the same qudit and same state (see for example circuit in~\figrefp{fig:1}{d}) can be contracted, reducing the number of blue sideband pulses. In this specific case, the pattern on the target \textit{qutrit} is then given by
\begin{equation}
	\sum_{j\in\{-1, 0, +1\}} c_j \ket{j} \mapsto e^{i3\Phi}c_{-1}\ket{-1} + e^{i2\Phi}c_{0}\ket{0} + e^{i3\Phi}c_{+1}\ket{+1},
\end{equation}
while the control qudit acquires $e^{i\Phi}$ on all spectator levels.

	We model the phase error $\Phi$ with a Gaussian distribution $\mathcal{N}(0,\sigma_{\text{p}}^2)$ with mean $\mu=0$ and variance $\sigma_{\text{p}}^2$.
	
\begin{figure}[ht]
		\centering
		\includegraphics[width=\columnwidth]{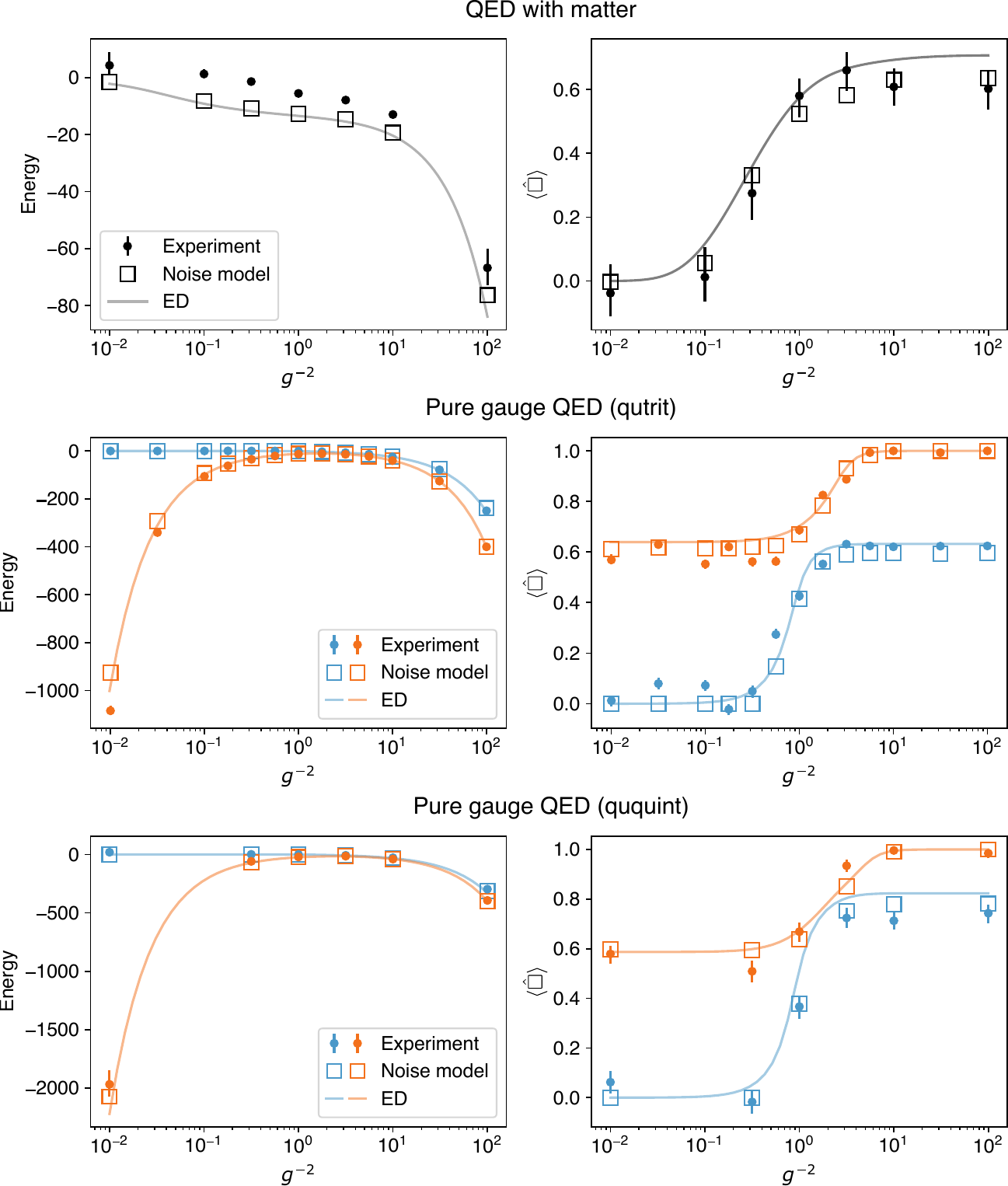}
		\caption{\textbf{Simulated noise model for the experiment.} Energy (first column) and plaquette expectation value (second column) for the two models studied via experimental VQE, namely 2D-QED with matter (first row), and an instance of pure gauge 2D-QED (second and third row, corresponding to qutrit and ququint truncations respectively). For the pure gauge model, blue (orange) lines and markers correspond to the simulations performed with the electric (magnetic) representation. After obtaining variational parameters in the absence of noise, the simulated noisy data (squares) is obtained via the noise model detailed in the text, which includes amplitude fluctuations and phase errors. The results from experimental quantum computations (dots) show good agreement with the simulated noisy data.}
		\label{fig:simulated_noise}
	\end{figure}

	\textbf{Results. } In \figref{fig:simulated_noise} we compare the experimental VQE results with simulated data for the open plaquette (first row) and for the pure gauge theory (second and third row). In the latter case the blue (orange) markers correspond to the electric (magnetic) representation. For the simulation, we consider variational parameters obtained without noise, and we calculate the expectation values of the relevant observables when the variational circuit is affected by the noise with parameters estimated as $\sigma_{\text{a}}^2= 0.0.047$, $\sigma_{\text{p}}^2=0.073$. We can see good qualitative agreement between the experimental and the simulated data.
	
	\section{Results for Alternative Lattice QED Model}
	\label{sec:model_cm}
	
	In this work, we consider compact lattice QED (cQED) with staggered fermions~\cite{kogut1975hamiltonian}. There are two different models that are both referred to as cQED and that are both currently discussed in the recent literature.
	The main text of this paper focuses on the first of these two models. As explained in Methods~\ref{sec:Methods_Hamiltonian}, this model is obtained by discretising space in two-dimensional QED~\cite{crippa_towards_2024, wiese2013ultracold}. In this section, we focus on the second version of cQED. This model has been studied in the literature on quantum simulations of lattice gauge theories~\cite{zohar_digital_2017, Magnifico_2021, paulson_simulating_2021, Clemente2022}, especially in condensed matter physics~\cite{bender_variational_2023, affleck_large-n_1988, rantner_electron_2001}. Both models respect a local U(1) gauge symmetry, but the kinetic parts of the Hamiltonian have different phases, which leads to different theories in the continuum limit.

	The Hamiltonian in the second case is given by~\cite{zohar_digital_2017}
	\begin{subequations}\label{eq:Hamiltonian_terms_cm}
		\begin{align}
			\hat{H}_E &= \frac{1}{2}\sum_{\vec{n}}\left(\hat{E}_{\vec{n},\vec{e}_x}^2 + \hat{E}_{\vec{n},\vec{e}_y}^2 \right), \\
			\hat{H}_B &= -\frac{1}{2}\sum_{\vec{n}}\left(\hat{P}_{\vec{n}} + \hat{P}_{\vec{n}}^\dagger \right), \\
			\hat{H}_m &= \sum_{\vec{n}} (-1)^{n_x+n_y} \hat{\phi}_\vec{n}^\dagger \hat{\phi}_{\vec{n}}, \\
			\hat{H}_k &= \sum_{\vec{n}} \sum_{\mu = x,y} \left(\hat{\phi}_\vec{n} \hat{U}_{\vec{n},\vec{e}_\mu}^\dagger \hat{\phi}_{\vec{n}+\vec{e}_\mu}^\dagger + \mathrm{H.c.}\right).
		\end{align}
	\end{subequations}
	
	Since the two models only differ by sign choice in the kinetic term (compare Eq.~(\ref{eq:openplaq_qubit}d)), they coincide for a pure gauge theory, as considered in Sec.~\ref{sec:periodic} of the main text. After eliminating the gauge fields as shown in Ref.~\cite{paulson_simulating_2021} and applying the Jordan-Wigner transformation given in Eq.~\eqref{eq:JW_transformation} the resulting qubit Hamiltonian is given by:
	
	\begin{subequations}
		\label{eq:openplaq_qubit_cm}
		\begin{align}
			\begin{split}
				\hat{H}_{E} &= \frac{1}{4} ( 8 \hat{E} + 2\hat{E}
				\left( - \hat{\sigma}^{z}_{1} - 2 \hat{\sigma}^{z}_{2} + 3 \hat{\sigma}^{z}_{3} + 2 \right) \\
				& \qquad \quad - \hat{\sigma}^{z}_{2} + \hat{\sigma}^{z}_{3} + \hat{\sigma}^{z}_{1} \hat{\sigma}^{z}_{2} + 3 ), 
			\end{split}\\
			\hat{H}_{B} &= -\frac{1}{2} \left( \hat{U} + \hat{U}^{\dagger}\right), \\
			\hat{H}_{m} &= \frac{1}{2} \left( \hat{\sigma}^{z}_{1} - \hat{\sigma}^{z}_{2} + \hat{\sigma}^{z}_{3} - \hat{\sigma}^{z}_{4} \right), \\
			\hat{H}_{k} &=  \hat{\sigma}^{+}_{1} \hat{\sigma}^{-}_{2} + \hat{\sigma}^{+}_{2} \hat{U}^{\dagger}\hat{\sigma}^{-}_{3}  + \hat{\sigma}^{+}_{4} \hat{\sigma}^{-}_{3}  - \hat{\sigma}^{+}_{1} \hat{\sigma}^{-}_{4}  + \text{H.c.},
		\end{align}
	\end{subequations}
	
\begin{figure}[ht]
		\centering
		\includegraphics[width=0.9\columnwidth]{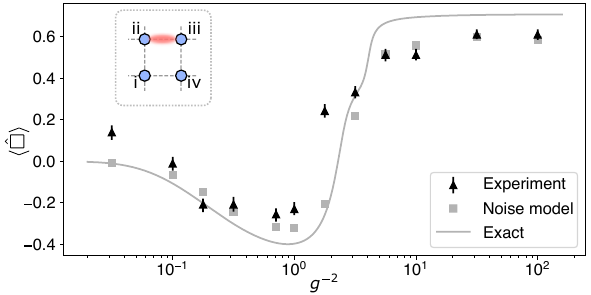}
		\caption{Experimental VQE results for the plaquette expectation value for the model described in Eqs.~\eqref{eq:openplaq_qubit_cm}.}
		\label{fig:open_plaquette_cm}
	\end{figure}

	We studied this model experimentally using the same VQE circuit structure as in \figrefp{fig:1}{d} (see Ref.~\cite{paulson_simulating_2021}). The gates that entangle the qubits and the qudit are still controlled on the qubits adjacent to the dynamical gauge field, which in this case are qubits ii and iii, as shown in the inset of \figref{fig:open_plaquette_cm}. The experimentally measured plaquette expectation values in the ground state found via VQE as a function of $g^{-2}$ are displayed in \figref{fig:open_plaquette_cm} for $\Omega=5$ and $m=0.1$, and show good agreement with the theoretical prediction. 
	
	In the large coupling regime ($g^{-2} \ll 1$) the electric field energy term $\hat{H}_E$ dominates the Hamiltonian of Eq.~\eqref{eq:openplaq_qubit_cm}, favoring the ground state $|vvvv,0\rangle$ with $\langle\plaqop\rangle= 0$. In the weak coupling regime, on the other hand, the magnetic field energy term $\hat{H}_B$ dominates the Hamiltonian, favoring a positive vacuum plaquette expectation value ($\langle\plaqop\rangle = \frac{1}{\sqrt{2}} \approx 0.707$ for the chosen truncation). In the intermediate regime where $g^{-2}\approx1$, we encounter a competition between the two field-energy terms and $\hat{H}_{k}$. The presence of the kinetic term $\hat{H}_{k}$ makes it energetically favorable to create pairs, which in turn produce a background magnetic flux that opposes the flux present in the ground state preferred by the term $\hat{H}_B$. On its own, the term $\hat{H}_{k}$ leads to a ground state with a negative plaquette expectation value ($\langle\plaqop\rangle \approx -0.642$ in the considered case). The presence of dynamical matter, therefore, results in a reduction of the plaquette expectation value in the ground state, as shown in \figref{fig:open_plaquette_cm}. Notably, even with our small system size we already clearly observe the dip in the plaquette expectation value that originates from pair creation processes in the ground state, see Ref.~\cite{paulson_simulating_2021} for details.
	
\end{document}